\documentclass[twocolumn]{aastex62}

\received{January 1, 2021}
\revised{January 25, 2021}
\accepted{\today}

\submitjournal{AJ}

\shorttitle{FUV Variable Sources in M31}
\shortauthors{Leahy et al.}
\begin{document}

\title{FUV Variable Sources in M31}

\correspondingauthor{Denis Leahy}
\email{leahy@ucalgary.ca}

\author{Denis Leahy}
\affiliation{Department of Physics and Astronomy, University of Calgary, Calgary, AB T2N 1N4, Canada}

\author{Megan Buick}
\affiliation{Department of Physics and Astronomy, University of Calgary, Calgary, AB T2N 1N4, Canada}

\author{Joseph Postma}
\affiliation{Department of Physics and Astronomy, University of Calgary, Calgary, AB T2N 1N4, Canada}

\author{Cole Morgan}
\affiliation{Department of Physics and Astronomy, University of Calgary, Calgary, AB T2N 1N4, Canada}

\begin{abstract}
The Ultraviolet Imaging Telescope (UVIT) onboard the AstroSat observatory has imaged the Andromeda Galaxy (M31) from 2017 to 2019 in FUV and NUV with the high spatial resolution of $\simeq1\arcsec$.
The survey covered the large sky area of M31 with a set of observations (Fields) each 28 arcminutes in diameter.
Field 1 was observed in two epochs with the F148W filter, separated by $\simeq$1133 days ($\simeq$3.10 years).
The 6.4 kpc diameter Field 1  (at the distance of M31) includes a substantial part of the inner spiral arms of the galaxy. 
We identify UVIT sources in both epochs of Field 1 and 
obtain catalogs of sources that are variable in FUV at $>3\sigma$ and $>5\sigma$ confidence level. 
The fraction of FUV variable sources is higher for brighter sources, and the fraction is higher in the two main spiral arms compared to 
other areas. 
This is evidence that a significant fraction of the FUV variables are associated with hot young stars.
Source counterparts are found for 42 of the 86 $>5\sigma$ FUV variables using existing catalogs. The counterparts include 10 star clusters, 6 HII regions, 5 regular or semiregular variables, 6 other variables and 6 nova or nova candidates.
The UVIT FUV-NUV and FUV-FUV color-magnitude diagrams confirm the association of most of the FUV variables with hot young stars.
A catalog of UVIT photometry for the variable sources is presented.
\end{abstract}


\section{Introduction} \label{sec:intro}

The Andromeda Galaxy (M31) is a large spiral with many similarities to our Galaxy and our closest neighboring giant spiral galaxy. 
It can therefore be used as a template to study parts of our Galaxy that are obscured because of the extinction between our position and theirs.
Another advantage to studying objects in M31 is that it has a known distance of 785 $\pm25$  kpc \citep{2005MNRAS.356..979M} . As a result, the uncertainty in intrinsic brightness for many objects in M31 is better known than for Galactic sources
as most Galactic sources have distance errors greater than those for M31 (3.2\%) .

At optical wavelengths, M31 has been observed on numerous occasions. The highest resolution observations are from the Hubble Space Telescope, including the Pan-chromatic Hubble Andromeda Treasury (PHAT) survey \citep{PHAT}. In near and far ultraviolet wavelengths (NUV and FUV), the GALEX instrument has surveyed M31 \citep{2005ApJ...619L...1M}. Recently M31 has been surveyed in NUV and FUV  \citep{2020ApJS..247...47L} with the UltraViolet Imaging 
Telescope (UVIT) on AstroSat. 

AstroSat has five instruments. The UltraViolet Imaging Telescope (UVIT) covers the FUV and NUV while the Soft X-ray Telescope (SXT), Large Area Proportional Counters (LAXPC), Cadmium-Zinc-Telluride Imager (CZTI) 
and Scanning Sky Monitor (SSM) 
instruments cover soft through hard X-rays \citep{2014SPIE.9144E..1SS}.
UVIT observations have high spatial resolution ($\simeq$1 arcsec) and the capability to resolve individual stellar clusters and stars in M31.
Previous UVIT observations of M31 are presented by 
\citet{2020ApJS..247...47L}, \citet{2020ApJS..250...23L}, \citet{2020arXiv201202727L} and \citet{2018AJ....156..269L}. These papers give a UVIT point source catalog for M31, a match of UVIT sources with Chandra sources in M31, matching UVIT sources with PHAT sources and analysis of UV bright stars in the bulge, respectively.

In this paper, we present results from two epochs of FUV observations in the central 28 arcmin of M31. FUV variable sources 
are found at significance levels of $>3\sigma$ (555 sources) and $>5\sigma$ (86 sources). 
The population frequencies of variable sources are compared for different regions. 
Color-magnitude diagrams (CMDs) of  variable and non-variable sources are compared, indicating that the variable sources are younger and more luminous on average.
Counterparts to $>5\sigma$ sources are identified by cross-matching with catalogs in the Vizier Web service, and their nature further investigated by NUV and FUV CMDs.  
The Appendix describes data processing that improves the photometric accuracy for moderately crowded regions.

\section{Observations and Data Analysis}

\begin{deluxetable*}{ccccccc}
\caption{M31 UVIT Observations for UVIT field 1.}
\centering
\tabletypesize{\footnotesize}
    \tablehead{\colhead{Observation} &         \colhead{RA(deg)$^{a}$} & \colhead{Dec(deg)$^{a}$} & \colhead{Filter$^{b}$} & \colhead{Exposure Time} & \colhead{Mean BJD$^{c}$}}
\startdata
    A     & 10.71071 & 41.25023 & a, c, d, e & 7872, 3875, 7920, 4347 & 2457671 (+0.8010,+1.2626,+0.8010,+1.2626) \\
    B     & 10.71071 & 41.25023 & a, b, c & 17191, 10427, 18177 & 2458804 (+1.3243,+0.8591,+2.2231) \\
\enddata
\tablecomments{a. RA and Dec are the J2000 coordinates of the nominal pointing center of the observation.\\
b. Filter labels are a: F148W, b: F169M, c: F172M, d: N219M, e: N279N.\\
c. Mean BJD is the mean solar-system Barycentric Julian Date of the observation. The common integer part for multiple observations is given as the first number.}
\label{table:obs} 
\end{deluxetable*}

\begin{figure*}[htbp]
    \centering
    \epsscale{1.1}
 \plottwo{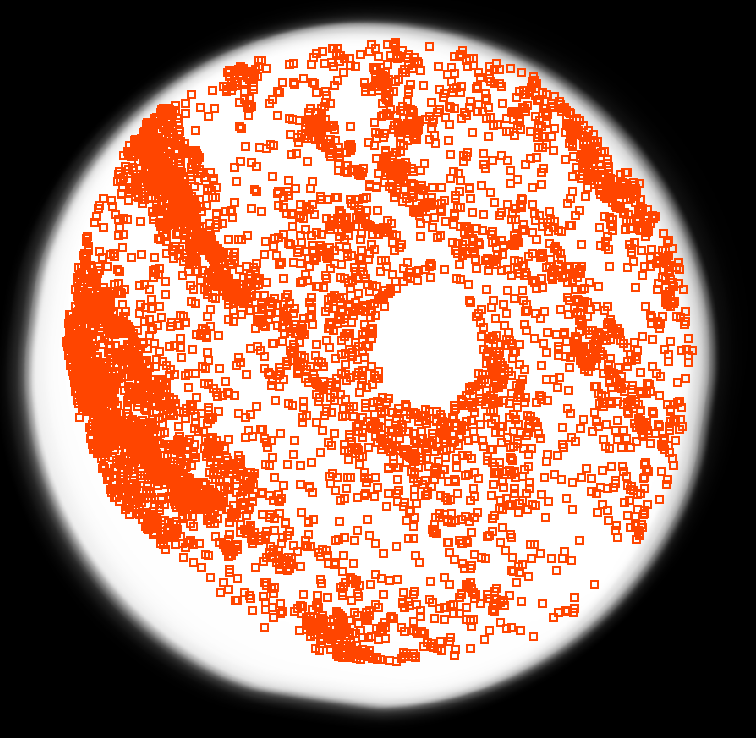}{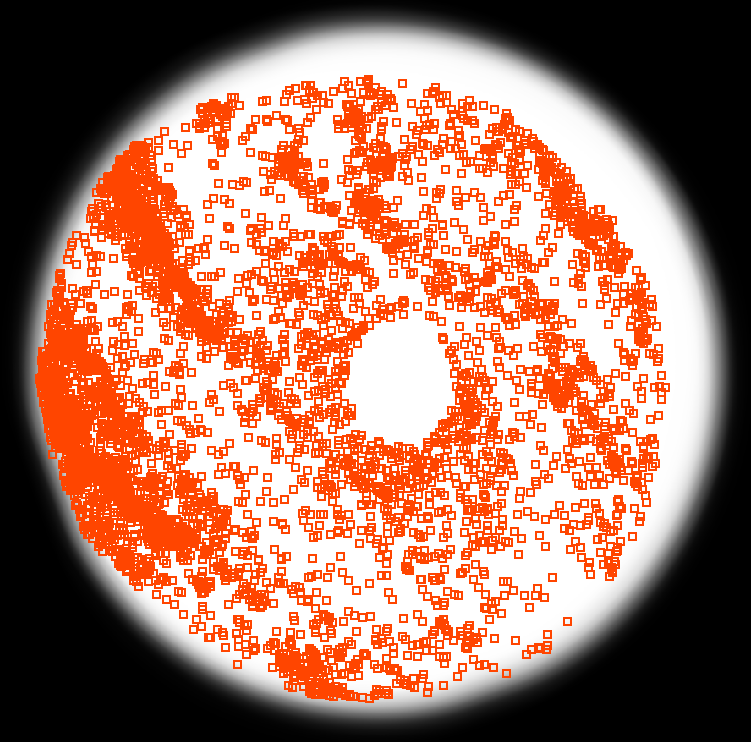}
    \label{fig:UVITexp}
    \caption{Field 1 F148W exposure maps for Observation A (left panel) and Observation B (right panel).
    The greyscale at bottom goes from black (0\% of exposure time) to white (100\% of  exposure time).
    The center of Observation B is shifted by $\simeq$2.2 arcminutes NW of the center of Observation A.
    The red squares mark the locations of the sources detected as non-variable in both images.
}
\end{figure*}

M31 was observed in FUV and NUV at $\sim1\arcsec$ spatial resolution by UVIT \citep{2020ApJS..247...47L}.
It was covered in nineteen 28 arcmin diameter fields (Fig. 2 of \citealt{2020ApJS..247...47L}).

UVIT Field 1 was observed in the F148W, F172M, N219M and N279N filters near the beginning of the observing project and in a new observation that has not been reported previously. 
The first observation is referred to as Observation A, the second as Observation B. 
Observation B was made in filters F148W, F169M and F172M.
The F172M filter is narrow with $\delta\lambda=12.5$ nm vs. 50 nm for F148W \citep{2017AJ....154..128T}),
thus has lower S/N than images in F148W.
F148W is used for the variability search because it gives high S/N. 
Only a fraction of the variables show up as variable above the 3$\sigma$ level in F172M, and all of those are detected variables in F148W.
Thus we detect variable sources by analysis of the F148W data.

The field centers, filters, exposure times and dates of the observations for Field 1 are shown in Table \ref{table:obs}. 
The basic data processing was carried out using updated 
data processing and calibration methods \citep{2020ApJS..247...47L}, 
including new astrometry corrections \citep{2020PASP..132e4503P}.

\begin{figure*}[htbp]
    \centering
    \epsscale{1.1}
 \plotone{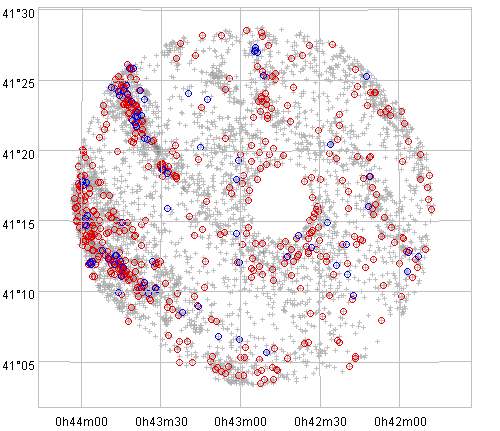}
    \label{fig:UVITposn}
    \caption{Field 1 with all identified UVIT sources from observations A and B (grey crosses), $>3\sigma$ variable sources (red circles) and $>5\sigma$ variable sources (blue circles). The central part of the bulge is visible as the blank area right of center.
}
\end{figure*}

\subsection{Source Finding}

Because of the satellite pointing drift of AstroSat, the edges of the UVIT field of view have low exposure time and
high noise level. Thus we avoid the outer edge for source finding.
We measured a position 
along a line NW from the centre 
in the exposure map where it has a value of 10\% for observation A. 
Then we took one position 0.5 arcmin closer to the center of the image and measured an exposure of  $\simeq$70\%, 
and a second position another 0.5 arcmin closer to the center, where we measured an exposure of 
 $\simeq$95\%. Nearly the same values were obtained from measurements of the exposure image for observation B. 
 Our search for sources was restricted to areas with exposure values $\gtrsim$50\% in both images. 
 Thus we avoid areas that are within $\sim$1 armin of the edge of either image for source finding. 
Figure 1 shows the exposure maps for Observations A and B and the non-variable\footnote{The variable
sources are a small fraction of all sources so inclusion of both sets does not significantly change the delineation of the 
edge of the areas studied.} sources detected in both images.
This illustrates the different pointing centers of the two observations and the choice of sky area used for source 
finding in both images.

Because of the bright diffuse FUV emission from the bulge, the source finding methods 
(\citealt{2017PASP..129k5002P}, \citealt{2020PASP..132e4503P}, Postma \& Leahy, 2021 accepted for J. Astrophys. Astr.),
did not produce reliable results in a region of $\simeq 2$ armin radius centered on the bulge. 
 In future work, we plan to develop a source finding method for UVIT data that works reliably in regions of bright diffuse emission.
 Thus, a circular area with radius 300 pixels\footnote{The UVIT pixels are $0.4168^{\prime\prime}$ by $0.4168^{\prime\prime}$ .} (2.1 arcmin) centered on the bulge was removed for the current analysis. 
 The center of the bulge is offset west of center in the images: the omitted region is visible in Figure 1.
 
 Source locations were found in observations A and B in the F148W filter. We used a threshold peak pixel signal to noise ratio (SN) of 3 for A and 4.5 for B and a 
 threshold integrated SN\footnote{This is the minimum ratio of (detected counts above background summed over pixels in the detection window) to (background counts summed over pixels in the window).}
  of 30 for A and 45 for B.
 The higher SN level for observation B was chosen to  approximately match the sensitivity levels because 
 observation B had $\sim$3 times the exposure time\footnote{Other methods were used for finding sources in A and B down to the same sensitivity limit. This included source finding on the image produced by merging data from observations A and B, which was unsuccessful due to the different exposure times of the two images. The chosen method produced the best results.}. 

\subsection{Source Magnitude Extraction}

The UVIT photometry is calibrated with respect to a Curve of Growth (COG) extracted with radius 95 
pixels ($40^{\prime\prime}$) \citep{Tandon}. 
This method is not applicable in practice when sources are separated by less than $40^{\prime\prime}$: 
 an alternate extraction method is necessary. 
For the UVIT images of M31, with their large number of sources, testing revealed that fitting pixels in a 9x9 box 
gave good results and minimal interference by neighboring sources for the majority of sources.

Because of various factors, including residual errors in the calibration of the optics and detector non-uniformities 
and satellite pointing drift corrections (which are different for each observation), each source in the image has a 
slightly different PSF. There are also variations in the detected image for each source from photon counting statistics. 
Thus unique point-spread function (PSF) fits are required to each source.
We have tested extensively that sources are fit well and consistently by using an elliptical Gaussian with variable parameters.

The count rates from elliptical Gaussian fits were corrected for the UVIT point-spread-function as described in Appendix A.
They were then converted to AB magnitudes using the calibration determined in \cite{Tandon}.
We describe the details of this process  in Appendix~\ref{App1}, including a correction to match UVIT N279N magnitudes to 
HST/PHAT \citep{PHAT} F275W magnitudes.

A source match was performed on the  F148W filter source lists from observations A and B using the TopCat catalog 
 software  \citep{2005ASPC..347...29T}.
 Overall, 3164 sources were found in both observations A and B. 
Upon  removing the matched sources from each the source lists from observation A and B,
we found 1383 sources in A that are not detected in B
 (`AnotB' sources), and 1431 sources in B that are not detected in A (`BnotA' sources). 
These latter two source lists were used to determine upper limits of `AnotB' sources for Observation B,
and upper limits in `BnotA' sources  for Observation A. 
The upper limits were found by carrying out forced photometry at the known source positions, i.e. elliptical Gaussian 
fits with normalization (source flux in counts/s) as the only free parameter. The upper limit was 
taken as the fit flux (or zero if the fit flux was negative) plus its uncertainty.  
The total number of distinct sources detected in either A or B is 5970.
The positions of the sources are shown in Figure 1.
\begin{figure}[htbp]
    \centering
    \epsscale{1.2}
    \plotone{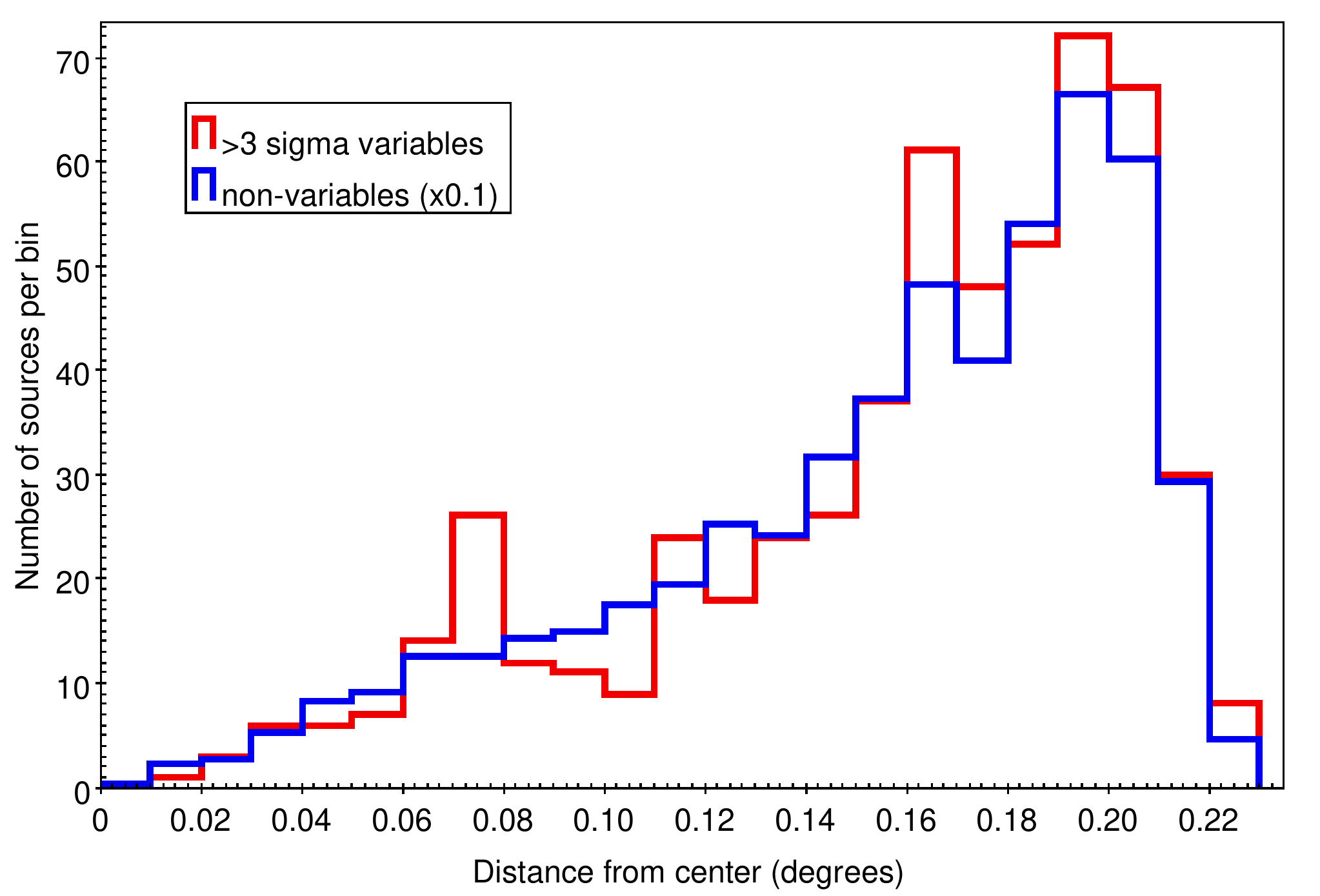}
    \label{fig:varnonvarVr}
    \caption{Distributions of variable and of non-variable sources vs. distance from the field center.}
\end{figure}

Using this list of F148W source positions, we carried out source finding on the N279N and N219M images from Observation A, and on the F172M and F169M images from Observation B. 
If the nearest light maxima in any of these filters was inconsistent with the F148W position (more than 1$\arcsec$ away), we list the filter magnitude as no detection. This yields UVIT multiband magnitudes for most of the sources. 

\begin{figure}[htbp]
    \centering
    \epsscale{1.2}
    \plotone{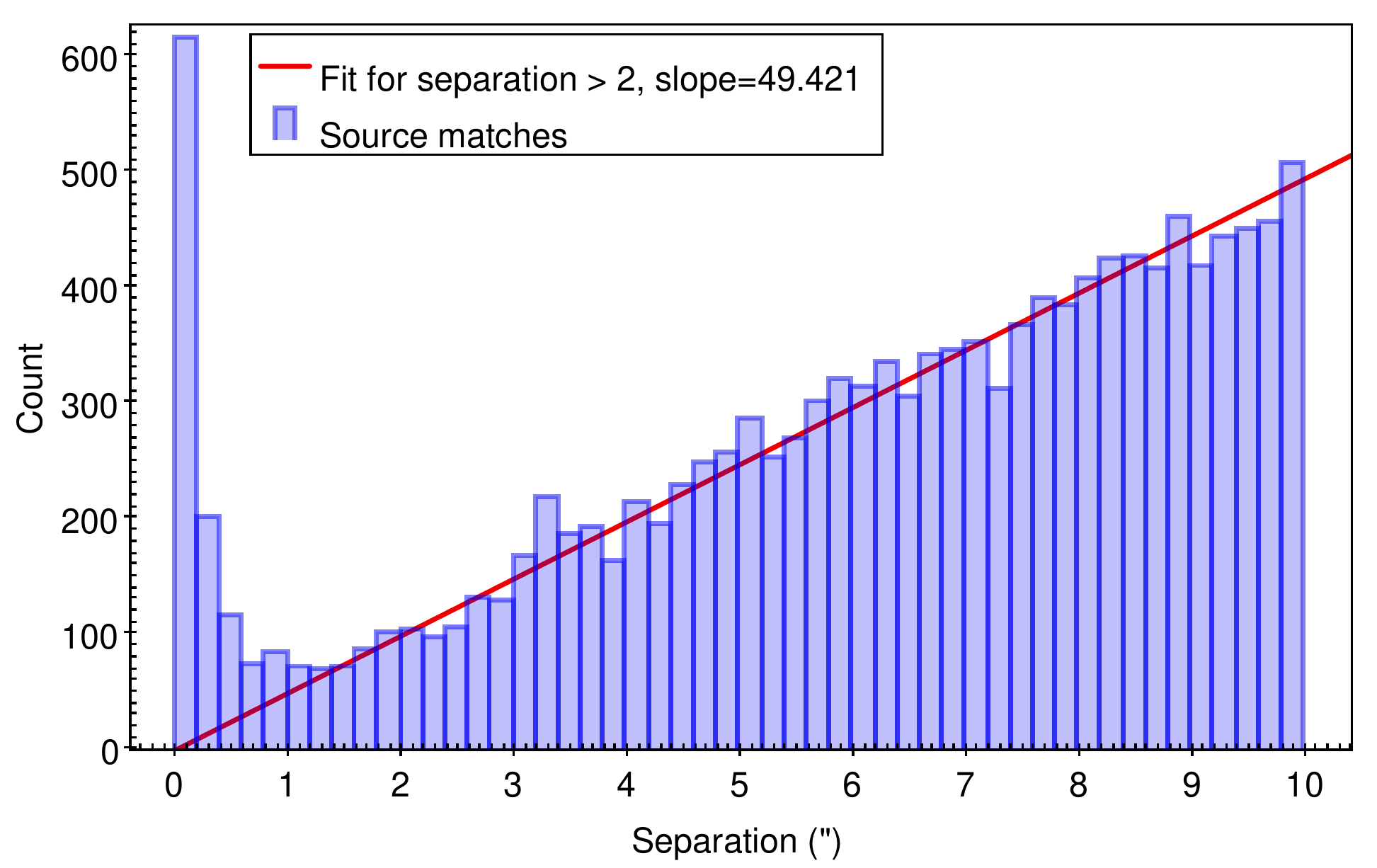}
    \label{fig:contam}
    \caption{Number of matches between sources in the UVIT Region 1 F148W at different separations. The line shows the best fit line to separations greater than 2$\arcsec$, which give the expected number of accidental matches.}
\end{figure}

\subsection{Identification of Variable Sources}

The result of the source finding and fitting was F148W magnitudes or upper limits for a total of 5970 different sources.
Requiring that the difference in magnitudes between observations A and B is $>3$ times the 
combined uncertainty of A and B magnitudes (labelled $\sigma$ here)
in the measurements, the number of $>3\sigma$ variable sources is 555. 
86 of these sources are variable at the $>5\sigma$ level\footnote{Using the source finding and fitting routines 
we originally detected 94 $>5\sigma$ variables. We manually inspected each fit and found 86 were good fits (true variables), 
3 were bad fits to non-variable sources and 5 were blended/crowded sources that could not be fit, resulting
in 86 verified $>5\sigma$ variables.}
The positions of the $>3\sigma$ variable sources and $>5\sigma$ variable sources are shown in Figure 2.

We checked that the variable sources are not detected preferentially near the field edges 
(see Fig. 1 for the field edges seen in
the exposure maps for Observations A and B). The radial distribution of source numbers vs. distance 
from the mean field center (average of field centers of Observations A and B) is shown in Figure 3.
The numbers of non-variable sources per bin are multiplied by a factor of 0.1 so that the distribution of variables
and non-variables can be compared easily. It is seen that they agree on the large scale, within Poisson counting
errors, with no peak at the  field edge. It is noted that the field edge here is outside the largest bin plotted because
we only carried out source finding for $\gtrsim$1 arcmin from the field edges. 

\subsection{Positional Errors for UVIT Sources}

The UVIT F148W source positions were checked against the source positions in PHAT
\citep{PHAT}
 and against source positions in Gaia DR2 (\citealt{2018A&A...616A...1G}, \citealt{2016A&A...595A...1G}). 
The complete source list from the combined observation A and B measurements was matched to the PHAT F275W filter catalog for sources in PHAT brighter than Vega magnitude 21 within a search radius of 1$\arcsec$, yielding 1571 matches. 
 From these sources, the mean difference (PHAT-UVIT position) in RA is 0.128$\arcsec$ with a standard deviation of 0.238$\arcsec$. 
 In DEC, the mean difference  (PHAT-UVIT position) is 0.0068$\arcsec$ and the standard deviation is 0.158$\arcsec$. 

 The source list from the combined observation A and B measurements was matched to the Gaia DR2 positions with search radius of 1$\arcsec$, yielding 974 matches. 
 The mean difference in RA (Gaia-UVIT position) was 0.038$\arcsec$ with a standard deviation of 0.322$\arcsec$. 
 The mean difference in DEC (Gaia-UVIT position) was 0.014$\arcsec$ with a standard deviation of 0.221$\arcsec$.
We use the Gaia DR2 offsets as the best estimate of the mean UVIT position offset and scatter. 
Both are significantly smaller than the UVIT spatial resolution of 1$\arcsec$, indicating that our astrometry calibration worked well.

\begin{longrotatetable}
\begin{deluxetable*}{ccccccccccccccccc}
\tablecaption{Photometry$^{a}$ for $>3\sigma$ variable FUV sources measured in UVIT Field 1. \label{tab:Catalog}}
\tablewidth{700pt}
\tabletypesize{\scriptsize}
    \tablehead{\colhead{UVIT} &  \colhead{UVIT} & \colhead{$\rm{F148W\_A}$} & \colhead{$\rm{F148W\_A}$} & \colhead{$\rm{F148W\_B}$} & \colhead{$\rm{F148W\_B}$} & \colhead{$\rm{F148W}$} & \colhead{$\rm{N279N\_A}$} & \colhead{$\rm{N279N\_A}$} & \colhead{$\rm{N219M\_A}$} & \colhead{$\rm{N219M\_A}$} & \colhead{$\rm{F172M\_B}$} & \colhead{$\rm{F172M\_B}$} & \colhead{$\rm{F169M\_B}$} & \colhead{$\rm{F169M\_B}$} \\ 
    \colhead{$\rm{RA^{b}}$} & \colhead{$\rm{DEC^{b}}$} & \colhead{mag} & \colhead{err} & \colhead{mag} & \colhead{err} & \colhead{Sigma} & \colhead{mag} & \colhead{err} & \colhead{mag} & \colhead{err} & \colhead{mag} & \colhead{err} & \colhead{mag} & \colhead{err} }
    
    \startdata
    10.487075 & 41.190926 & 20.44 & 0.08 & 24.79 & 0.24 & 20.8 & 18.75 & 0.07 & 22.38 & 0.23 & . & . & . & . \\
10.925487 & 41.431048 & 20.09 & 0.08 & 22.12 & 0.1 & 18.06 & . & . & . & . & 19.76 & 0.06 & 19.78 & 0.05 \\
10.723582 & 41.449845 & 20.11 & 0.08 & 18.9 & 0.07 & 17.55 & 20.56 & 0.15 & . & . & . & . & . & . \\
10.992857 & 41.29603 & 21.92 & 0.12 & 19.78 & 0.07 & 16.95 & 19.48 & 0.09 & 19.5 & 0.07 & 19.37 & 0.05 & 19.55 & 0.05 \\
10.57196 & 41.162304 & 21.18 & 0.1 & 99.99 & 99.99 & 15.49 & 18.53 & 0.06 & 19.49 & 0.07 & . & . & . & . \\
10.922499 & 41.392021 & 22 & 0.12 & 20.13 & 0.08 & 13.63 & 20.12 & 0.13 & 19.65 & 0.07 & 19.73 & 0.06 & 19.97 & 0.05 \\
10.895092 & 41.165119 & 19.06 & 0.07 & 18.45 & 0.07 & 13.23 & 18.63 & 0.07 & 18.89 & 0.06 & 18.71 & 0.05 & 18.98 & 0.04 \\
10.723116 & 41.449416 & 19.88 & 0.08 & 19.04 & 0.07 & 12.91 & . & . & . & . & 21.13 & 0.11 & 22.19 & 0.13 \\
10.756415 & 41.299913 & 24.65 & 0.33 & 20.81 & 0.08 & 12.23 & . & . & . & . & 19.9 & 0.07 & 20.39 & 0.06 \\
10.880259 & 41.169574 & 21.94 & 0.12 & 20.29 & 0.08 & 12.07 & 20.18 & 0.13 & 20.28 & 0.09 & 19.93 & 0.07 & 20.29 & 0.06 \\
    \enddata
\tablecomments{a: Magnitudes are in the AB system; the value . indicates no detection. b: RA and DEC are measured in degrees.}
\tablecomments{Table 1 is published in its entirety in the machine-readable format.
      A portion is shown here for guidance regarding its form and content.}
\end{deluxetable*}
\end{longrotatetable}

The observed density of detected UVIT sources determines the number of accidental matches we can expect caused by source crowding.
The combined UVIT observation A and B  F148W sources were internally crossmatched up to a radius of 10$\arcsec$, keeping all matches for each source. 
The resulting distribution of numbers of sources vs. separation is shown in Figure 4. 
The sharp peak centered on 0$\arcsec$ separation indicates the real matches.  
Fitting the numbers with separations greater than 2$\arcsec$  with a linear regression line gave a slope of 49.4 counts/bin/$\arcsec$ (using bins of 0.2$\arcsec$). 

To estimate the contamination of accidental matches, we integrated the corresponding area under the line for sources between 0$\arcsec$ and 1$\arcsec$ to get 123. 
The number below the line for sources between 1$\arcsec$ and 2$\arcsec$ is 371.
To estimate the number of real matches we sum the actual sources above the line to get: 593 between 0$\arcsec$ and 1$\arcsec$ and 29 between 1$\arcsec$ and 2$\arcsec$.
This means we are getting approximately 123 accidental matches between 0$\arcsec$ and 1$\arcsec$ (about 17\% of the matches in  0-1$\arcsec$), and we are missing about 29 real matches between 1$\arcsec$ and 2$\arcsec$ (about 7\% of the matches in  1-2$\arcsec$) by choosing a 1$\arcsec$ upper limit for UVIT matches.
If we chose a smaller cutoff, we could reduce the accidentals at the expense of losing more real sources, but we take 1$\arcsec$ as a realistic compromise.

\section{Results and Discussion}

\subsection{Catalog of FUV Variables}

For observation A, the UVIT filters were F148W, N219M and N279N, allowing multiband photometry for the sources detected in the F148W filter.  
Similarly for observation B, the UVIT filters were F148W, F169M and F172M.
From this set of photometry, we created a catalog of the $>3\sigma$ variables, which includes the $>5\sigma$ variables.
This catalog is presented here (first 10 entries) by Table~\ref{tab:Catalog}, and in full as an online machine readable table.

\subsection{Quantifying the Variability}
\begin{figure}[htbp]
    \centering
    \epsscale{1.1}
    \plotone{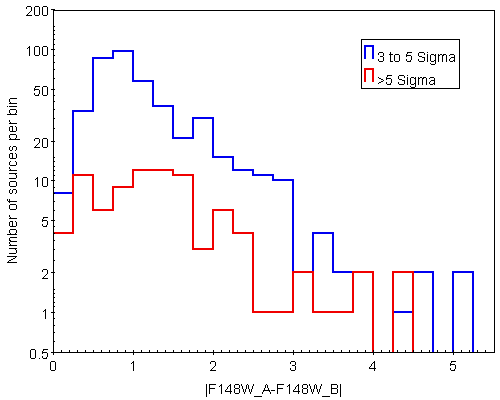}
    \label{fig:magdiff}
    \caption{Histogram of F148W magnitude differences, for $>5\sigma$ set and $3$ to $5\sigma$ set (log scale on vertical axis). 
    The mean magnitude difference for the $>5\sigma$ set is 1.45, for the $3$ to $5\sigma$ set is 1.23.}
\end{figure}

\begin{figure*}[htbp]
    \centering
    \epsscale{1.1}
    \plottwo{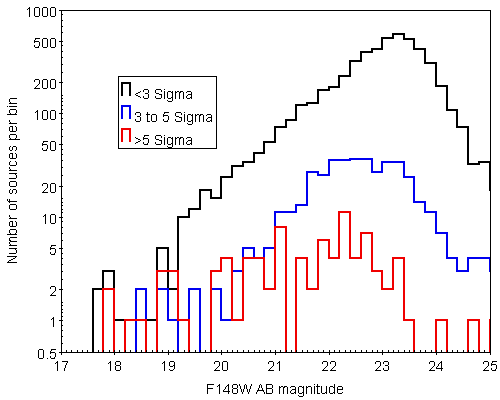}{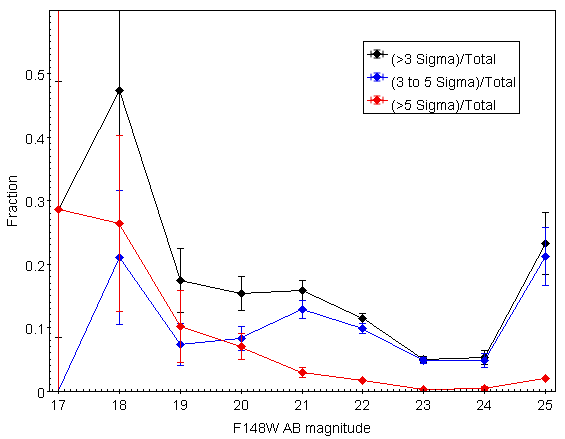}
    \label{fig:mag_hist}
    \caption{Left panel: Histogram of magnitudes for variable and non-variable sources, using F148W magnitudes from Observation A.  
    Right panel: Fraction of variables in different categories to total number of sources, in 1 magnitude wide bins.}
\end{figure*}

The magnitude differences between the two observations, separated by $\sim3.1$ years, were found for the set of 3 to $5\sigma$ variables and for the set of  $>5\sigma$ variables.  Figure 5 
shows the histograms of magnitude differences for both sets. 
The mean magnitude differences translate to a mean flux ratio of 3.1 for the 3 to $5\sigma$ variables, and a mean flux ratio of 3.8 for the $>5\sigma$ variables.
As seen in Figure 5, 
the drop in number of sources for magnitude differences $\lesssim0.75$ for 3 to $5\sigma$ variables, and  for  $\lesssim0.25$ for $>5\sigma$ variables can be explained by the typical magnitude errors in the UVIT measurements 
($\simeq0.06$ at F148W magnitude 17, increasing smoothly to $\simeq0.2$ at magnitude 23).

The distribution of magnitudes from observation A for the variable and non-variable sources is shown in the left panel of Figure 6. 
The majority of sources are non-variable ($<3\sigma$), with the turnover at F148W magnitude of $\simeq$23.4 representing the approximate sensitivity limit of UVIT F148W detection, for an observation time of 8000s. 
The turnovers in the distributions of 3 to $5\sigma$ variables and  $>5\sigma$ variables are at $\simeq$23 and $\simeq$22.6, respectively.
These are consistent with observational limits determined by magnitude errors.
Above these limits, the shape of the magnitude distribution should reflect the intrinsic distribution of source numbers.
Of interest is the fraction of variable sources as a function of magnitude. This is illustrated in right panel Figure 6.
The error bars are calculated using counting errors and increase for bright sources because of the few sources in the brighter magnitude bins.
The decrease in fraction of variables for faint sources ($\gtrsim22$) is due to observational limits.
The increase in fraction of variables for bright sources should be intrinsic. 
The variable fraction for 3 to $5\sigma$ variables is $\sim15$\%, and is consistent with no change as a function of magnitude.
For the $>5\sigma$ variables, the fraction of variable sources shows weak evidence for an increase with increasing brightness.

\subsection{Spatial Distribution of the Variables}

\begin{figure}[htbp]
    \centering
    \epsscale{1.1}
 \plotone{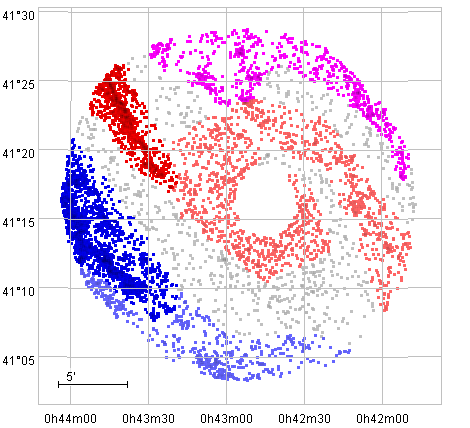}
    \label{fig:3groups}
    \caption{Areas used for spatial analysis of variable sources.
    All UVIT variable and non-variable sources from observations A and B are shown by the squares (grey and colored); 
    The sources in the dense part of the southeast arm (`SEarm') are shown in blue; those in the dense 
    part of the northeast spiral arm (`NEarm') in red; those in the extended part of the southeast arm 
 (`Ext.SEarm') by light blue; those in the extended part of the northeast arm (`Ext.NEarm') by pink;  those 
 in the extended part of the northwest arm (`Ext.NWarm') by magenta; and those not in either the dense or extended parts of the arms (`interarm') by grey.    
    The central part of the bulge is visible as the blank area left of center.
}
\end{figure}

\begin{table*}[htbp]
  \centering
  \caption{Numbers of sources, $>3\sigma$ variables and $>5\sigma$ variables in different areas, and their ratios.}
    \begin{tabular}{cccccccc}
    \hline
    \hline
    & SEarm & NEarm & NotSE,NE$^1$  & Ext.$^1$SEarm & Ext.$^1$NEarm&  Ext.$^1$NWarm &  interarm$^1$ \\
    \hline
    total number & 938   & 1704  & 3330  & 1364   & 538  & 768 & 660 \\
  $>3\sigma$ var. & 101   & 187   & 268  & 118   & 50   & 48 & 52 \\
   $>5\sigma$ var. & 23    & 30    & 35  & 15   & 5   & 8  & 7 \\
    $>3\sigma$ var./total  & $0.108 \pm 0.011$ & $0.110 \pm 0.008$  & $0.081 \pm 0.005$  & $0.087 \pm 0.008$ & $0.093 \pm 0.014$  & $0.063 \pm 0.009$ & $0.079 \pm 0.011$ \\
  $>5\sigma$ var./total  & $0.025 \pm 0.005$ & $0.018 \pm 0.003$ & $0.011 \pm 0.002$ & $0.011 \pm 0.003$ & $0.09 \pm 0.004$ & $0.010 \pm 0.004$ & $0.011 \pm 0.004$ \\
    \hline
    \end{tabular}
  \label{tab:Distribution}
	  \tablecomments{1.  NotSE,NE stands for everthing but the areas covered by SEarm and NEarm;  
	  Ext. stands for Extended, and does not include the SEarm or NEarm areas; 
	  the interarm area shown by the grey symbols in Figure 5.}
\end{table*}

 A visual comparison of all UVIT sources with the $>3\sigma$ variable sources in Figure 2 
 shows that the variable sources appear to be concentrated in the two main spiral arms in the southeast and northeast parts of the image.
 To test this, we used the polygon feature in TopCat to select a number of different groups of sources. 
 The seven groups consisted of one group each for the dense part of the southeast and northeast spiral arms (SEarm
 and NEarm); 
 a group in the extended part of the southeast arm (Ext.SEarm); 
 a group in the extended part of the northeast arm (Ext.NEarm); 
 a group in the extended part of the northwest arm (Ext.SWarm); 
  a group of sources including all other sources except those in the southeast and northeast spiral arms (notSE,NE); 
 and a group for interarm sources (not in either the dense parts of the arms or the extended parts of the arms: notSE,NE, notExt.SE,NE,SW).
 These first 5 of these groups are shown by the color-coded symbols in Figure 7 and the last group shown by the grey symbols. 

Table~\ref{tab:Distribution} shows the results from counting sources in the different groups with Poisson errors. 
The results show the SEarm and NEarm have the same fraction of $>3\sigma$ variables within errors. 
The area notSE,NE  has a lower fraction of $>3\sigma$ variables compared to the NEarm, with 3$\sigma$ confidence.
A comparison of the Ext.SE,  Ext.NE and  Ext.NW arms shows that the fraction of $>3\sigma$ variables is consistent between the  
Ext.SE and  Ext.NE areas, but higher than the  Ext.NW area by 2$\sigma$ confidence. 
The difference between the fraction of $>3\sigma$ variables in Ext.SEarm or Ext.NEarm and the interarm area is 
only $\sim1\sigma$ confidence.

For the 5$\sigma$ variables, the SEarm has a higher fraction of variables than the NEarm by 
1.2$\sigma$ confidence, and higher by 2.6$\sigma$ confidence than for the NotSE,NE area. 
The extended arm areas and the interarm area are consistent with each other in fraction of variables.
In summary, we find that the dense arms (SEarm, NEarm) have a significantly larger fraction of variables than the rest of the field, considering both $>3\sigma$ and $>5\sigma$ variables. 
However there is no significant difference between the extended arms and the interarm area. 

The UVIT F148W filter (150 nm band) is sensitive to detection of hot luminous stars. The above results imply that the fraction of variable sources is significantly higher in the regions with the highest hot star density compared to either the regions with lowest density or medium density. 
The regions with the lowest density and with a medium density of hot stars have indistinguishable fractions of variable stars.
In particular the high density SEarm and NEarm have higher fraction of variables by a factor of $\sim$1.4 (from the 
$>3\sigma$ variables in Table~\ref{tab:Distribution}) to $\sim$2 (from the $>5\sigma$ variables) compared to the other 5 
areas listed in the Table.

\subsection{Colour Magnitude Diagrams}

\begin{figure}[htbp]
    \centering
    \epsscale{1.2}
    \plotone{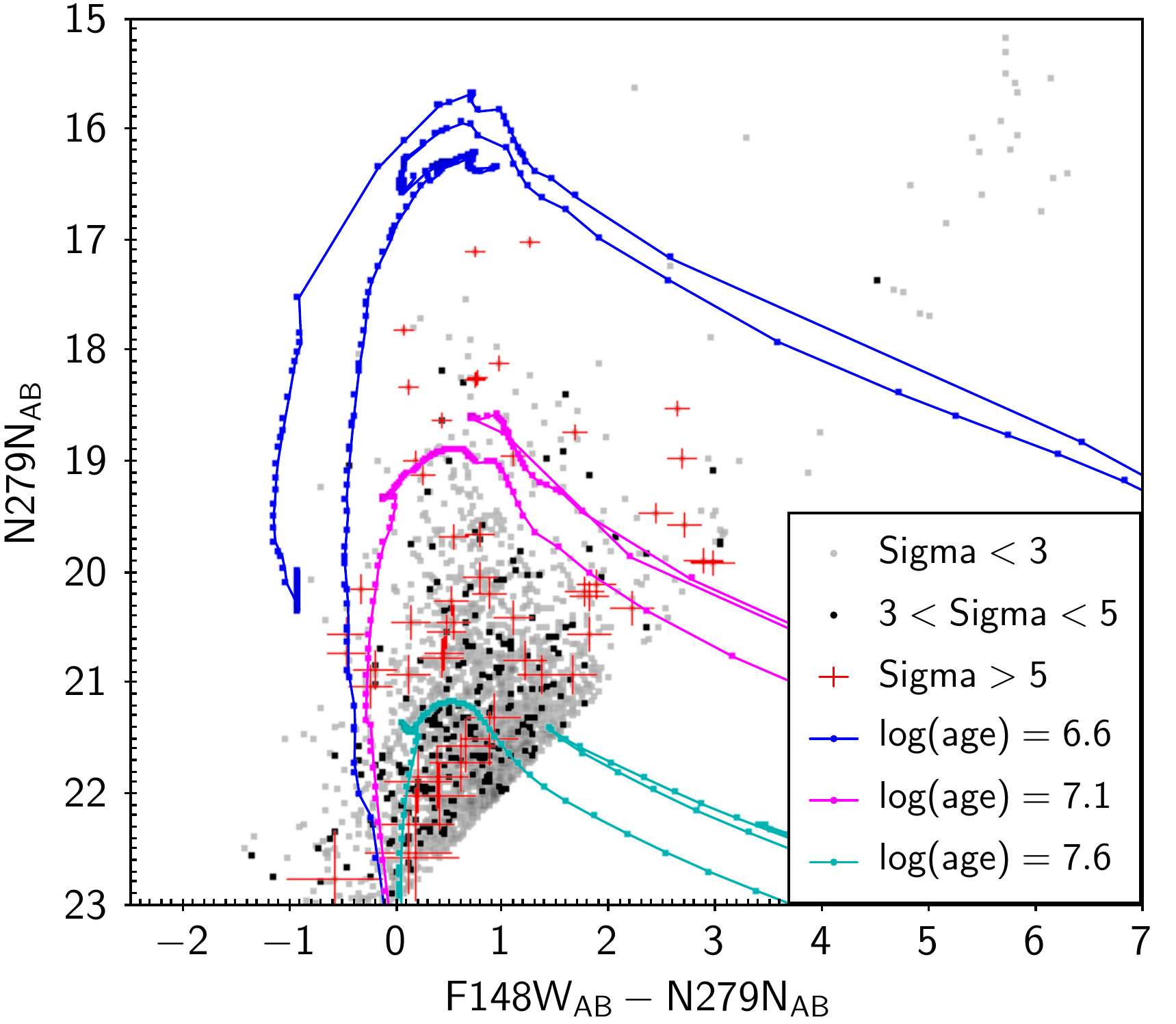}
    \label{fig:ObsA_CMD}
    \caption{Field 1 Observation A colour magnitude diagram (CMD):  N279N versus F148W-N279N. 
     The gray points are all detected sources in Observation A, 
    the black points are the $>3\sigma$ variables, and the red points with error bars are the $>5\sigma$ variables.
    The error bars for the gray and black points are similar in size, but not shown to avoid crowding.
    Isochrones are from the CMD web isochrone generator at http://stev.oapd.inaf.it/cmd, and are plotted as the points joined by solid lines, with log(age) values of 6.6 (blue lines),
    7.1 (magenta lines) and 7.6 (cyan lines).
 }
\end{figure}

\begin{figure}[htbp]
    \centering
    \epsscale{1.2}
    \plotone{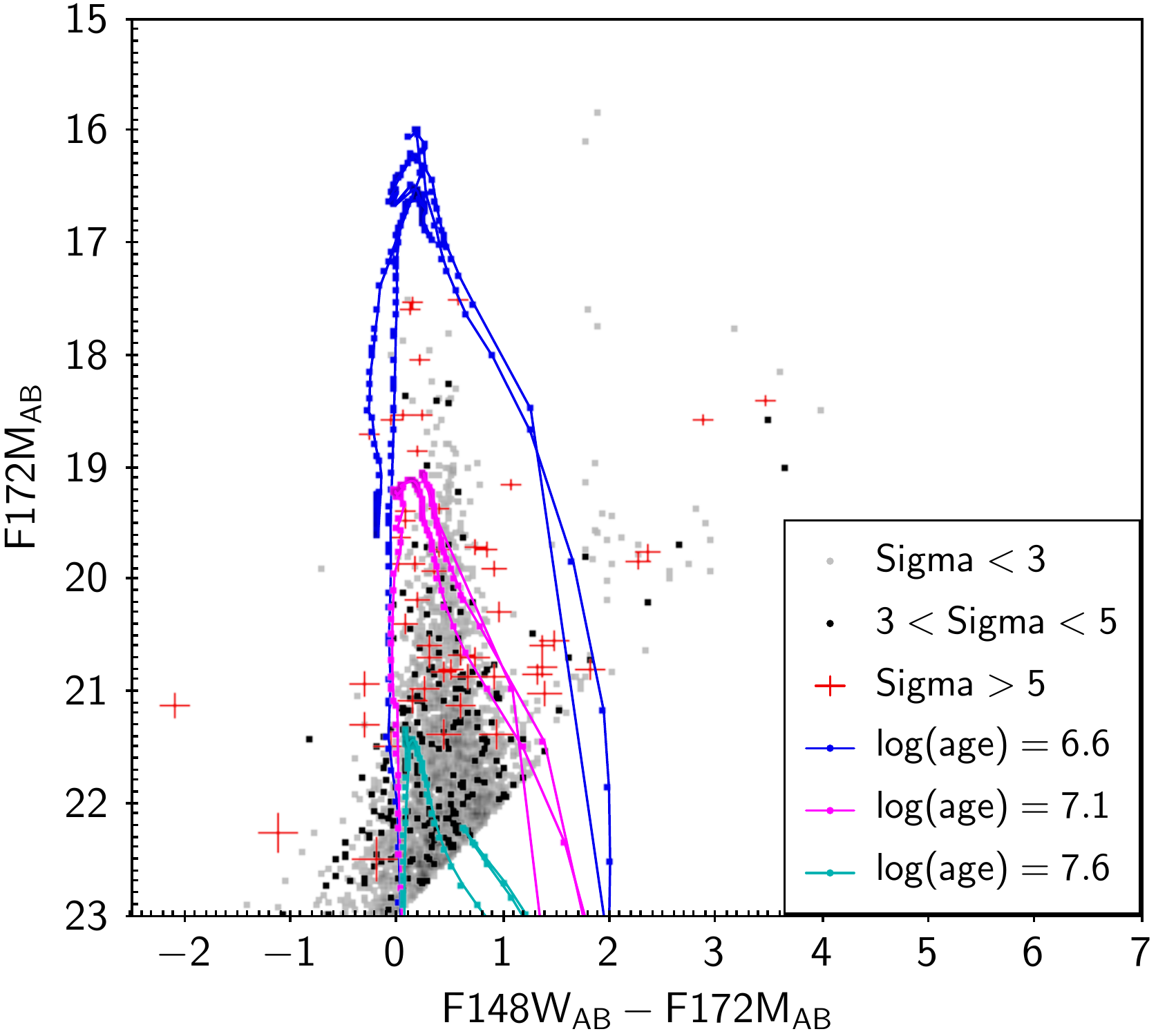}
    \label{fig:ObsB_CMD}
    \caption{Field 1 Observation B colour magnitude diagram (CMD): F172M versus F148W-F172M.
     The gray points are all detected sources in Observation B, 
    the black points are the $>3\sigma$ variables, and the red points with error bars are the $>5\sigma$ variables.
    The error bars for the gray and black points are similar in size, but not shown to avoid crowding.
    Isochrones are from the CMD web isochrone generator at http://stev.oapd.inaf.it/cmd, and are plotted as the points joined by solid lines, with log(age) values of 6.6 (blue lines),
    7.1 (magenta lines) and 7.6 (cyan lines).
}
\end{figure}

From the photometry for the variable sources and the photometry for the non-variable sources ($<3\sigma$ difference in F148W magnitudes between Observation A and Observation B), we construct FUV-NUV CMDs.
For Observation A, we use FUV-NUV color and NUV AB magnitude, with NUV chosen as either N279N or N219M, for the three sets: all detected sources; $3-5\sigma$ variables; and $>5\sigma$ variables.
The  N279N vs. F148W-N279N CMD is shown in Figure 8. 
The  N219M vs. F148W-N219M CMD looks similar.
For Observaton B, there were three FUV filters, allowing construction of CMDs using FUV 2-color and FUV AB magnitude, for the same three groups of sources.
The F172M vs. F148W-F172M CMD is shown in Figure 9. 
The F169M vs. F148W-F169M CMD looks similar.

On the CMDs, we plot isochrones from the CMD web isochrone generator at http://stev.oapd.inaf.it/cmd. 
These isochrones use solar metallicity, the foreground extinction to M31 of $A_V$=0.2 and other parameters set to default in the isochrone generator. 
We also calculated isochrones for metallicity of $log(Z/Z_{\odot})$=-1 and -2 but this did not result in a large shift of the isochrones 
(comparable to the size of the data error bars). 
Increased extinction moves the isochrones almost vertically downwards (increase in magnitude but little change in color). 

From Figures 8 and 9, 
the majority of the data points are consistent with stars aged $\sim5\times10^6$ yr to $\sim1\times10^8$ yr. 
These are young stars. The UVIT observations select stars which are brightest in FUV and NUV bands, so we expect a strong selection for the youngest stars in M31.
A small fraction of the data points lie at the upper right in the diagram, away from where the model isochrones lie. 
These points are probably cool foreground stars in the Milky Way. If those stars were at the distance of M31 they would be 
shifted down by the difference in distance modulus (e.g., by 14.46  magnitudes if the foreground stars are at 1 kpc). 
This would move them into the region consistent with isochrones with ages of a few Gyr (at N279N magnitude of $\sim$30).

\subsection{Source Matching to Counterparts at Other Wavelengths}

For the set of 86 $>5\sigma$ variable sources, a search was performed for previously detected objects. 
The counterpart search was carried out using the online tool Vizier (website https://vizier.u-strasbg.fr/viz-bin/VizieR, 
\citealt{2000A&AS..143...23O}).	
The UVIT positions are accurate to better than 1$\arcsec$ (1$\sigma$ error of 0.2$\arcsec$) but the position accuracy of the counterparts is different for each catalog, typically $\sim1\arcsec$, so we chose to use a search radius of 2$\arcsec$ within the UVIT position of each source.
Several of the counterparts are extended sources (stellar clusters). Thus for those, the cluster effective radius, $R_{eff}$, is given in Table~\ref{tab:Vizier}. The $R_{eff}$ values range from 0.4 to 2.5 $\arcsec$. This was a factor in the choice of  
the 2$\arcsec$ search radius, because we did not want to miss stellar clusters within which the UVIT source resides.

The results are shown in Table~\ref{tab:Vizier}, with counterpart position, type and distance in $\arcsec$ from the UVIT position listed.
One UVIT variable source had 3 counterparts, which are highly likely to be a single object with the reported positions differing by 
up to 0.1$\arcsec$.
10 FUV variables had 2 listed counterparts, and the remainder had  1  counterpart.
For the 10 cases with 2 counterparts (rows 2 to 11 in the table), most are likely the same object. 
The exceptions are row 2, where the globular cluster is not likely to be a counterpart, and row 9, where the 
two objects are very different and we chose to put the planetary nebula first because of its much better position match.
Some cases are ambiguous, e.g. for the sixth line in the table, the UVIT variable is within the cluster but because of the larger position offset may or may not be associated with the Wolf-Rayet star.
 
The following types of counterparts for the UVIT FUV variables were found. 
The main categories were: 9 matches with the \cite{Johnson} cluster catalog; 6 matches with ionized hydrogen (HII) regions; 6 matches with nova or nova candidates; 6 matches with unspecified variables and 5 matches with regular or semiregular variables.
Smaller categories included: 2 hot supergiants and 1 S Doradus variable; 2 eclipsing binaries; 1 planetary nebula (PNe); 1 Cepheid; 1 Supernova Remnant (SNR) candidate; 1 unspecified transient and 1 foreground star in the Gaia DR2 catalog. 
One of the sources matching the cluster catalog also matched with a Wolf-Rayet star.
\begin{figure*}[htbp]
    \centering
    \epsscale{1.1}
    \plotone{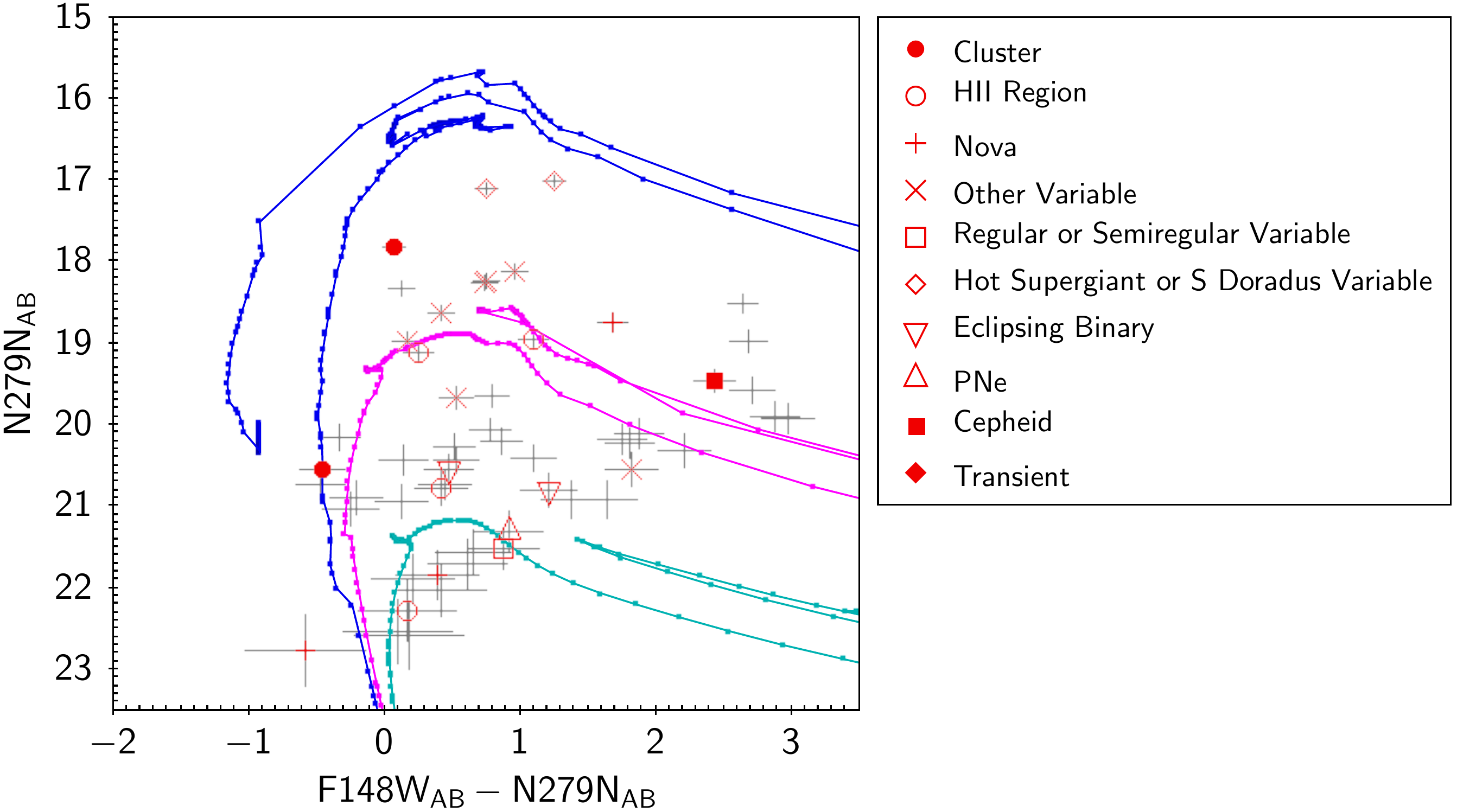}
  \plotone{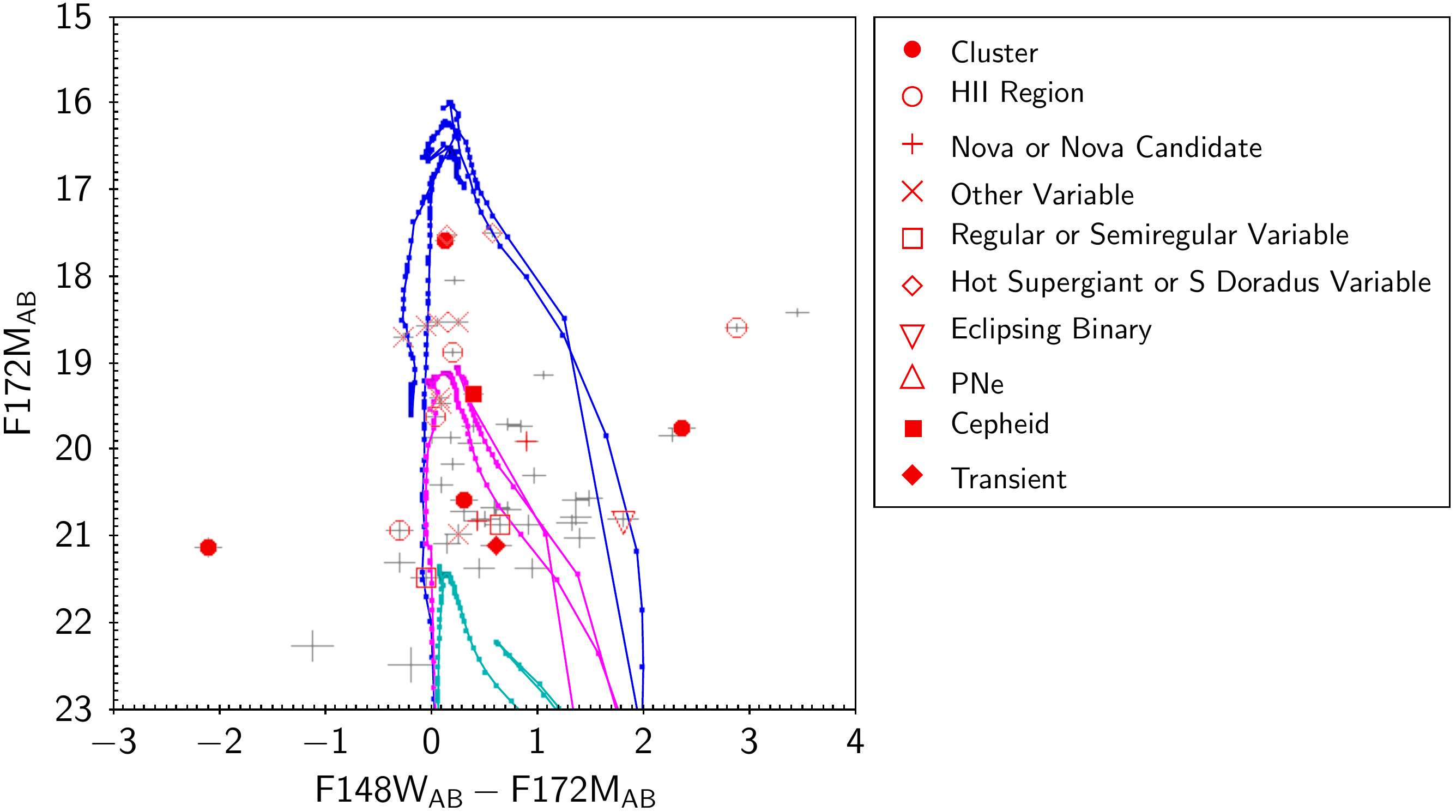}
    \label{fig:ObsAB_CMD2}
    \caption{Top panel: Field 1 Observation A colour magnitude diagram (CMD) with N279N versus F148W-N279N. 
    The gray points with error bars are the FUV $>5\sigma$ variable source measurements from Observation A, 
    and the red symbols mark the identified counterparts from Table~\ref{tab:Vizier}. 
    Bottom panel: Field 1 Observation B colour magnitude diagram (CMD) with F172M versus F148W-F172M. 
  In this case, the gray points with error bars are the FUV $>5\sigma$ variable source measurements from Observation B.
    Isochrones are from the CMD web isochrone generator at http://stev.oapd.inaf.it/cmd, and are plotted as the points joined by solid lines, with log(age) values of 6.6 (blue lines), 7.1 (magenta lines) and 7.6 (cyan lines).
 }
\end{figure*}

Overall, 42 of the 86 UVIT FUV $>5\sigma$ variables matched previously identified sources at other wavelengths.
The clusters and HII regions are generally associated with young hot stars.
Thus, most of the 42 identified sources are matched with a variable or match with a phenomenon associated with variability.
The current results based on variability in FUV are therefore consistent with previous surveys.

The UVIT CMDs for the FUV $>5\sigma$ variables is shown in Figure 10  
with Observation A data in the top panel and Observation B data in the bottom. 
The counterpart types are labelled using the First Catalog Match from Table~\ref{tab:Vizier}. 
Not all 42 counterparts appear in both diagrams because a given counterpart may not have been measured in enough filters, or have a 
bright enough F148W, F172M or F179N in the given observation to show up on the CMD.

The brightest FUV variables, above the $10^{7.1}$ year isochrone in both panels of Figure 10 
are those identified as hot supergiants or residing in clusters, which is consistent with their locations in the CMDs.
The S Dor star and the two hot supergiants have F148W magnitudes in Observations (A, B) of   (18.24, 18.06), 
(17.83, 17.66) and (21.7, 22.9), respectively. Because the two brighter stars are so bright in FUV, the relatively small FUV variability of
$\sim0.2$ magnitudes is significant at $>5\sigma$. 
The FUV variables intermediate in luminosity, near the $10^{7.1}$ year isochrone, are identified
with some HII regions, other variables, and the Cepheid variable. 
The fainter FUV variables, near the the $10^{7.6}$ year isochrone, are identifiied with the eclipsing binaries, the remaining HII regions, regular or semiregular variables, one cluster and the PNe.   
The UVIT source identified with the cluster, Source No. 15  in Table~\ref{tab:Vizier},  is the object that is the outlier in the lower panel of Figure 10 with F148W-F172M of $-2.1$.
This UVIT source is at the edge of the cluster, 1.86$\arcsec$ from the cluster center, so may be an object with a bright UV 
flare unrelated to the cluster.
The other object with a large change in color is Source No. 24, associated with a Cepheid variable. 
The F148W magnitude in changes from 21.89 in Observation A to  19.75 in Observation B.

Detailed modelling of the individual sources is beyond the scope of this work. The UVIT FUV variable source photometry
catalog is published here to enable further investigation of individual sources. 
The full catalog is available online and a sample (the first 10 lines) of the catalog is given here in Table~\ref{tab:Catalog}.

\section{Summary and Conclusion}

Using a new observation of M31, Field 1 of the M31 UVIT survey now has been observed in FUV at two epochs separated by $\simeq$3.1 years.
Field 1 is a 28 arcminute diameter (6.4 kpc) circular region centered on the M31 inner spiral arms and bulge. 
Using the source finding algorithm in CCDLAB, a list of source positions was constructed.
 These positions were then matched against the PHAT and Gaia DR2 catalogs. 
 The analysis of differences in RA and in DEC and their standard deviation 
 yields an astrometric accuracy of $\sim0.2\arcsec$  ($1\sigma$) for the UVIT data.
We carried out an analysis of the photometry of extracted sources (details given in  Appendix~\ref{App1}),
which allows for calibrated photometry in M31 in regions of moderate source crowding.

Field 1 in M31 includes a significant area around the bulge showing clear spiral arm structure (e.g. see images in
Figures 1 and 2 of \citealt{2018AJ....156..269L}). 
The two epochs of Field 1 imaging in the UVIT F148W filter allows detection of variable FUV sources. 
A total 3164 sources were detected in both Observations A and B, 1383 were only detected in A, and 1431 were only detected in B. 
 From this total of 5970 sources, 555 ($\simeq$9\%) were found to be variable by $>3\sigma$ and 
86 ($\simeq$1.5\%) were found variable by $>5\sigma$. 

The mean FUV flux ratio from the two observations was found to be 3.1 for the  3 to 5 $\sigma$ variables, and 3.8 for the $>5\sigma$ variables. 
The magnitude distributions 3 to 5 $\sigma$ variables and $>5\sigma$ variables are flatter than for non-variables,  well above the UVIT detection limit, giving evidence for a larger fraction of variables for the bright sources than for fainter sources.
Analysis of the spatial distribution shows that a greater number of the variables are found in the spiral arms than between the spiral arms
and that variables made up a greater percentage of the total number of sources in the spiral arms than outside the spiral arms.
This is consistent with most variables being associated with young stellar systems concentrated in the spiral arms.

\begin{longrotatetable}
\begin{deluxetable*}{ccccccccc}
\tablecaption{Multiwavelength Counterparts of the UVIT FUV $>5\sigma$ Variables$^{A}$. \label{tab:Vizier}}
\tablewidth{700pt}
\tabletypesize{\scriptsize}
\tablehead{  \colhead{Source No.} &  \colhead{ra} & \colhead{dec} & \colhead{First catalog match} & \colhead{Distance} & 
    \colhead{$\rm{R_{eff}}$} & \colhead{Second/third catalog match} & \colhead{Distance} 
}
\startdata
 1 &   10.8878 & 41.2029 & S Doradus variable$^{1}$ & 0.18  & & Luminous blue variable$^{13}$ & 0.02  \\
    &  &  & & & &   No coherent periodicity to variability$^{15}$ & 0.06  \\
  2 &  10.9133 & 41.1721 & Hot supergiant variable$^{2}$ & 0.85  &   & Globular cluster$^{14,19}$ & 0.69    \\
   3 & 10.7224 & 41.4502 & Hot supergiant variable$^{2}$ & 0.32  &       & Dubious variable$^{15}$ & 0.17   \\
   4 & 10.8951 & 41.1651 & Cluster$^{3}$ & 1.223 & 2.56  & Variable$^{16}$ & 0.697   \\
   5 & 10.9456 & 41.2100 & Variable$^{4}$ & 1.092 &       & Ionized nebulae$^{17}$ & 0.748   \\
   6 &  10.9462 & 41.2107 & Cluster$^{3}$ & 0.941 & 2.30   & Wolf Rayet star$^{18}$ & 1.14    \\
   7 & 10.8020 & 41.3939 & Eclipsing binary$^{5}$ & 0.08  &       & Variable$^{4}$ & 0.761  \\
   8 & 10.8152 & 41.1498 & Nova$^{1}$ & 0.35  &       & Variable$^{4}$ & 0.286   \\
  9 &  10.6131 & 41.2485 & Planetary nebula$^{6}$ & 0.25  &   & Regular or semi regular variable$^{7}$ & 1.344   \\
 10 &   10.7564 & 41.2358 & Regular or semi regular variable$^{7}$ & 1.589 &      & Variable$^{4}$ & 1.175  \\
 11 &   10.5834 & 41.2227 & Regular or semi regular variable$^{7}$ & 1.887 &      & Variable$^{4}$ & 1.526   \\
 12 &   10.8734 & 41.3117 & Cluster$^{3}$ & 0.104 & 0.64  &       &         \\
  13 &  10.9463 & 41.2010 & Variable$^{4}$ & 1.305 &       &       &        \\
  14 &  10.7236 & 41.4499 & Cluster$^{3}$ & 1.001 & 2.08  &       &         \\
  15 &  10.7231 & 41.4494 & Cluster$^{3}$ & 1.86  & 2.08  &       &         \\
  16 &  10.8837 & 41.1705 & Variable$^{4}$ & 1.517 &       &       &       \\
  17 &  10.8715 & 41.3116 & Variable$^{4}$ & 0.656 &       &       &         \\
   18 & 10.8654 & 41.3066 & Cluster$^{3}$ & 0.335 & 0.41  &       &        \\
  19 &  10.9478 & 41.1987 & HII region$^{8}$ & 1.50   &       &       &        \\
 20 &   10.9972 & 41.2927 & HII region$^{8}$ & 0.20   &       &       &        \\
  21 &  10.9340 & 41.4046 & Variable$^{4}$ & 0.643 &       &       &        \\
 22 &   10.5480 & 41.2676 & Regular or semi regular variable$^{7}$ & 0.844 &       &       &      \\
 23 &   10.8645 & 41.2649 & HII region$^{8}$ & 0.40   &       &       &       \\
 24 &   10.9929 & 41.2960 & Cepheid variable$^{9}$ & 1.613 &       &       &      \\
 25 &   10.6581 & 41.2338 & Regular or semi regular variable$^{7}$ & 0.386 &       &       &       \\
 26 &   10.9331 & 41.1952 & Foreground source$^{10}$ & 1.373 &       &       &      \\
  27 &  10.9191 & 41.4306 & Cluster$^{3}$ & 0.247 & 0.44  &       &        \\
  28 &  10.7521 & 41.2013 & Regular or semi regular variable$^{7}$ & 1.656 &       &       &      \\
  29 &  10.9933 & 41.2283 & Cluster$^{3}$ & 0.079 & 0.88  &       &      \\
  30 &  10.9255 & 41.4311 & Cluster$^{3}$ & 0.269 & 0.49  &       &      \\
  31 &  10.9418 & 41.4034 & Eclipsing binary$^{5}$ & 0.23  &       &       &       \\
  32 &  10.9959 & 41.2458 & Variable$^{11}$ & 0.59  &       &       &     \\
  33 &  10.9879 & 41.1993 & HII region$^{8}$  & 0.33  &       &       &       \\
   34 & 10.8643 & 41.3113 & HII region$^{8}$ & 1.08  &       &       &       \\
  35 &  10.4871 & 41.1909 & Nova$^{12}$  & 0.92  &       &       &        \\
  36 &  10.8137 & 41.3379 & Nova Candidate$^{20}$ & 0.68  &       &       &       \\
  37 &  10.9535 & 41.4080 & HII region$^{8}$  & 1.73  &       &       &      \\
  38 &  10.5813 & 41.1871 & Supernova Candidate$^{21}$ & 0.03  &       &       &     \\
  39 &  10.7564 & 41.2999 & Nova$^{22}$ & 0.24  &       &       &        \\
  40 &  10.7536 & 41.3218 & Nova$^{23}$ & 1.11  &       &       &       \\
  41 &  10.6379 & 41.2191 & Nova Candidate$^{24}$ & 0.10   &       &       &     \\
  42 &  10.5981 & 41.1979 & Transient event$^{25}$ & 0.22  &       &       &       \\
\enddata
\tablecomments{ A:  ra and dec are sky position of the UVIT FUV variable in decimal degrees; first, second, and third 
catalog matches refer to the identified typing of the object in catalogs in Vizier; distance refers to 
the separation in arcseconds between the UVIT source and the identified match from Vizier. 
$\rm{R_{eff}}$ gives the radius of the cluster from the literature for sources that match with a cluster.\\ \\
Catalog references: 1. \cite{Samus}, 2. \cite{HumphreysIV}, 3. \cite{Johnson}, 4. \cite{An}, 5. \cite{Vilardell}, 
6. \cite{Halliday}, 7. \cite{Fliri}, 8. \cite{Azimlu}, 9. \cite{Kodric}, 10. \cite{GaiaDR2}, 11. \cite{Kurtev}, 
12. \cite{HornochAug}, 13. \cite{HumphreysV}, 14. \cite{Kim}, 15. \cite{Heinze}, 16. \cite{Bonanos}, 
17. \cite{Walterbos}, 18. \cite{Neugent}, 19. \cite{Peacock}, 20. \cite{Williams2015}, 
21. \cite{HornochMar}, 22.\cite{HornochOct}, 23. \cite{HornochJul}, 24.\cite{Ovcharov}, 
25. \cite{Soraisam}.}
\end{deluxetable*}
\end{longrotatetable}

CMDs for Observation A and for Observation B 
 show that most sources detected in FUV are consistent with relatively young stellar populations.
The range of inferred age for the variables is $\sim5\times10^6$ yr to $\sim1\times10^8$ yr. 
The variable stars, in comparison to the non-variable stars are concentrated at younger ages, confirming that most variable stars are associated with younger stellar populations.
This is consistent with the results from the spatial analysis.

The $>5\sigma$ variables were position-matched  to catalogs of known sources. Of the 86 sources,
42 had a match to a previously known source. Most of these were with known variable sources.
The main categories include 
star clusters,  HII regions, nova or nova candidates,  regular or semiregular variables,  other variables, and hot supergiants.
This confirms that UVIT FUV measurements can select hot variable stars as well as other types of FUV variables such as novae.

A catalog of the UVIT photometry of the variables is presented, and is available on-line with this paper. 
The aim is to facilitate future multiwavelength research of individual objects of interest.

\acknowledgments
This work is supported by funding from the Canadian Space Agency. 
This publication uses data from the AstroSat mission of the Indian Space Research Institute
(ISRO), archived at the Indian Space Science Data Center (ISSDC).
This work has made use of data from the European Space Agency (ESA) mission
{\it Gaia} (\url{https://www.cosmos.esa.int/gaia}), processed by the {\it Gaia}
Data Processing and Analysis Consortium (DPAC,
\url{https://www.cosmos.esa.int/web/gaia/dpac/consortium}). Funding for the DPAC
has been provided by national institutions, in particular the institutions
participating in the {\it Gaia} Multilateral Agreement.
We thank the anonymous referee for comments which lead to several improvements in this article.

\appendix

\section{Tests to Determine Source Extraction for Crowded Fields\label{App1}}

The images we used were created using CCDLAB, including various corrections such as pointing drift correction,
flat field correction and astrometry calibration.
For one of the first tests, we compared source fitting on the CCDLAB image with source fitting on the image after smoothing. 
Convolution of the image with Gaussians of various FWHM
were tried, and we found that a the image smoothed with 1.5 pixel FWHM Gaussian
worked well as a compromise between producing better fits and not compromising spatial resolution.  This 1.5 pixel convolution is termed the smoothed image. 
The smoothed image resulted in better fits to point sources than the unsmoothed image, mainly
because of a much smaller fraction of bad fits (fits which failed to produce reasonable fits to the image).
As an example, the brightest 100 sources in N279N Field 2 were fit using smoothed and unsmoothed images.
The mean least-squares of the fits for the smoothed image was much smaller (2.6 vs. 6250 for the unsmoothed image).
 Because of these tests, the resultant convolution was applied to all of the images before fitting.

\begin{figure*}
\gridline{\fig{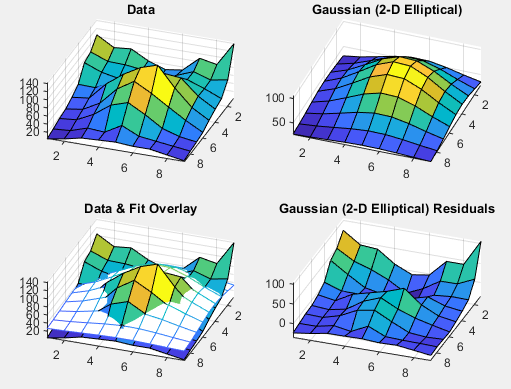}{0.5\textwidth}{(a)}
          \fig{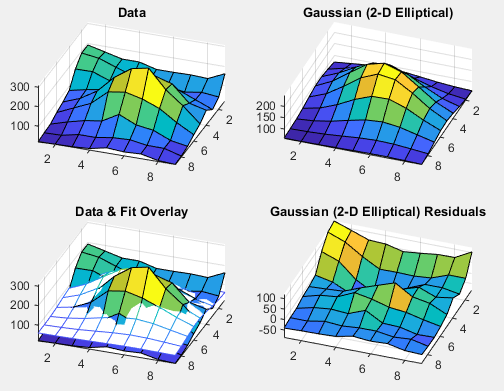}{0.5\textwidth}{(b)}}
          \gridline{\fig{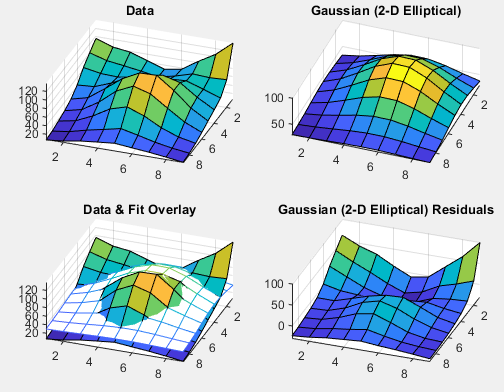}{0.5\textwidth}{(c)}
          \fig{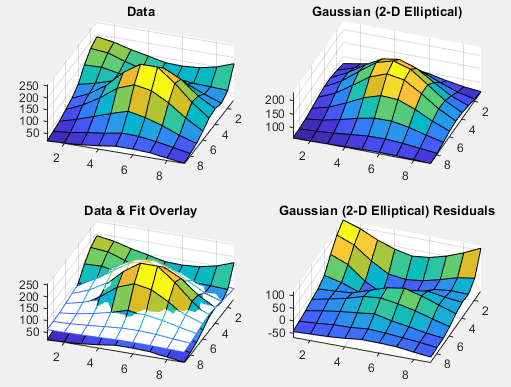}{0.5\textwidth}{(d)}}
\gridline{\fig{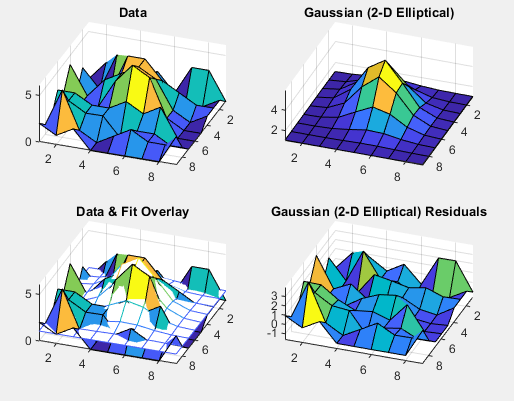}{0.5\textwidth}{(e)}
          \fig{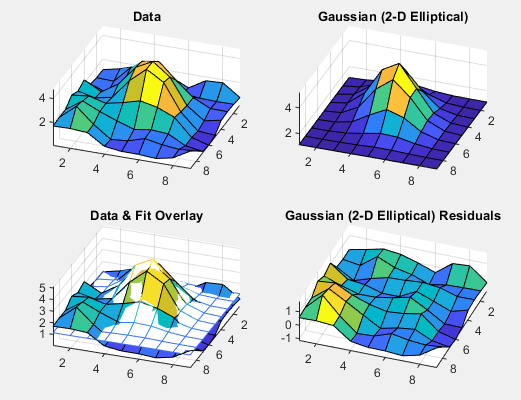}{0.5\textwidth}{(f)}}
\caption{Plot of Elliptical Gaussian fits to selected variable sources, showing data, model, data and model overlay, and residuals for each fit:
(a) Source N1 observation A (no smoothing); (b)  Source N1 observation B  (no smoothing); (c)  Source N1 observation A (smoothed);  (d)   Source N1 observation B  (smoothed); (e) Source SWbulge1 observation A  (no smoothing);
(f) Source SWbulge1 observation A  (smoothed); (see text for details). \label{fig:6plots}}
\end{figure*}

The HST-PHAT survey has accurate positions and photometry in the NUV F275W filter (275nm band), which we could use to calibrate the UVIT observations.
We used M31 Fields 1, 2, 7 and 13 for developing the source extraction method because they had N279N 280 nm band observations that overlap with HST-PHAT observations.
Field 2 was the main test field because it had the largest overlap with the HST-PHAT survey.

For Fields with N279N, 7x7 pixel circular and elliptical Gaussian and Moffat functions were fit to 15 bright, isolated sources for each filter. 
Similarly, Curves of Growth (COGs) with a half-width of 26 pixels were fit to each of these sources.
A linear regression line was fit to the COG fluxes vs Gaussian or Moffat fluxes for the 15 sources for each filter.
The fits gave conversion factors from flux for each of the other fitting types to the COG flux.
The resulting factors and their standard deviations are shown in Table \ref{tab:moff_gauss_COG}.

\begin{table}[htbp]
  \centering
  \caption{Conversion factors between 7x7 Circular and Elliptical Gaussian and Moffat fits and a COG with radius 26 for 15 bright sources in Field 2.}
    \begin{tabular}{ll}
    \hline \hline
          & Conversion Factor \\
    \hline
    Gaussian, Circular & 1.617$\pm$0.044 \\
    Gaussian, Elliptical & 1.580$\pm$0.037 \\
    Moffat, Circular & 0.460$\pm$0.716 \\
    Moffat, Elliptical & 0.386$\pm$0.396 \\
    \hline
    \end{tabular}
  \label{tab:moff_gauss_COG}
\end{table}

We examined the regression fits to determine which fitting type best fits a large number of sources with a small box size.
 The Moffat fits were frequently skewed by low volume sources, which it over-fit with unrealistic numbers. 
 On the other hand, the Gaussian fits were found to be proportional to the COG with a small standard deviation. 
 Since the elliptical Gaussian fits have the least variance when compared to the COG, this was chosen as the best fitting function to determine volumes. 

 To determine the best sized fitting box to use, 7x7 and 9x9 elliptical Gaussian fits were then performed on the N279N Field 2 image and compared with the F275W filter photometry from PHAT. 
 We also tried 5x5 but that resulted in many poor fits and incorrect photometry, and 11x11 but it had too many regions affected by adjacent sources.
The ratios in source counts from the Gaussian fit with 7x7 and 9x9 boxes to counts from the COG were found. 
The scatter (standard deviation) in the ratios was significantly larger for the 7x7 fits than for the 9x9 fits.
Thus 9x9 was chosen as the better size.
Figure 11 illustrates fits for an average-brightness source (top 4 panels) and a faint source (lower 
2 panels) from the F148W $>5\sigma$ variable source list.
Each panel shows data (upper-left sub-plot),
fit (upper right subplot), overlay (lower-left subplot) and residuals (lower-right subplot).
The top row (a and b) shows the fits for a source, labelled N1, in the northern part of the field shown in 
Figure 2, comparing the A (left) and B (right) source fits on the unsmoothed images.
The middle row (c and d) shows the fits for N1 on the smoothed A and B images.
The residuals are significantly lower for the smoothed images.
 This source has nearby sources at 2 of the 4 corners of the 9x9 region, thus it 
 would be impossible to use COG photometry on it. 
The bottom row (e and f) shows the fits for a faint source southwest of the bulge (southwest
of the blank area in Figure 2), labelled SWbulge1,
comparing the unsmoothed A (left) and smoothed A (right) fits.
The SWbulge source is faint and in a crowded region of high background, with probable fainter sources at 3 of the 4 corners of the 9x9 fit region. For this source the difference in residuals between smoothed and unsmoothed fits is apparent.

 To find the best constraints for the sources, the 100 brightest PHAT sources were fit with elliptical Gaussians in CCDLAB 
 and the range of parameters was constrained to the smallest and largest reasonable values. 
 Outliers (fits that failed to converge) included a few sources with sigmas less than 0.6 pixels and some 
 with sigmas greater than 4 pixels. Ignoring these, the sigmas of the fitted elliptical Gaussians 
ranged from 1.1 to 2.2 pixels, so these were chosen as constraints for the sources. 
Using the constraints, the least-squares of the fits was much lower than without the constraints.

 Using the smoothed N279N Field 2 image, constrained 9x9 elliptical Gaussians were fit and compared with COGs of half width 12, 14, 17, 20 and 25. 
 At a width of 25, nearby sources were visible at the edge of the fit, so this was the largest COG used. 
 For the other Fields and filters there were not enough isolated sources unless we used only the COG half widths of 12, 14 and 17. 
 The COGs at each of these radii correspond well with the trend observed in \cite{Tandon}. 
 Thus, a first conversion factor was found from the 9x9 elliptical Gaussian to each of the 12, 14, and 17 half-width COGs,
 and a second conversion factor from these smaller COGs to a 95 pixel COG using the table from \cite{Tandon}.
 An overall conversion factor was then obtained from combining the two conversion factors for COG half-width, then
 averaging the values from the different COG half-widths. 
 This was carried out for  each filter. 
 A list of factors for the different Fields and filters is shown in Table \ref{tab:conv_fac}.

\begin{table*}[htbp]
  \centering
  \caption{Gaussian to COG Conversion factors for M31 Fields 1, 2, 7, and 13 in UVIT.}
    \begin{tabular}{cccccc}
    \hline
    \hline
   Field & F148W  &  F172M & F169M & N279N & N219M \\
    \hline
    1     & $1.529\pm 0.093$ & $1.653\pm 0.014$ & $1.542\pm 0.017$ &  $1.654\pm 0.020$ & $1.621\pm 0.025$ \\
    2     & $1.694\pm0.027$ & $1.650\pm0.043$ & N/A & $1.646 \pm 0.009$ & $1.646\pm0.028$ \\
    7     & $1.606\pm 0.035$ & $1.647\pm 0.017$ & N/A & $1.784\pm 0.012$ & $1.942\pm 0.020$ \\
    13    & $1.671\pm 0.016$ & $1.582\pm 0.008$ & N/A & $1.719\pm 0.021$ & $1.820\pm 0.019$ \\
    \hline
    \end{tabular}
  \label{tab:conv_fac}
\end{table*}

 Using the trapezoidal integral function in numpy in python (numpy.trapz), the Vega flux and standard AB flux were 
 integrated over the N279N filter profile. From this, the difference in magnitude between the two measurement systems 
 was found to be 1.483. This allows for comparison between the PHAT and UVIT magnitudes \citep{PHAT}.

 The corrected magnitude in N279N was compared against the F275W PHAT magnitudes for sources brighter than 
 AB magnitude 22 in UVIT and brighter than Vega magnitude 20.5 in PHAT. 
 The two were fit with the python function scipy.optimize.curve\_fit to determine the difference in intercept between the two. 
 From this, a difference of $1.613\pm0.013$ was obtained. 
As such, an additional correction factor of $0.130$ was applied to the magnitudes, which corresponds to the 
difference between the measured and expected difference between UVIT and PHAT. 
 These correction factors are given in Table \ref{tab:Correc}.

\begin{table}[htbp]
  \centering
  \caption{Correction for N279N in UVIT based on F275W in PHAT.}
    \begin{tabular}{cc}
    \hline
    \hline
    Field & Correction (magnitudes) \\
    \hline
    1     & $0.105\pm0.021$ \\
     2     & $0.130\pm0.013$ \\
    7     & $0.168\pm0.013$ \\
    13   &  $0.075\pm 0.031$ \\
    \hline
    \end{tabular}
  \label{tab:Correc}
\end{table}


\begin{thebibliography}{}

\bibitem[{{An} {et~al.}(2004){An}, {Evans}, {Hewett}, {Baillon}, {Novati},
  {Carr}, {Creze}, {Giraud-Heraud}, {Gould}, {Jetzer}, {Kaplan}, {Kerins},
  {Paulin-Henriksson}, {Smartt}, {Stalin}, \& {Tsapras}}]{An}
{An}, J.~H., {Evans}, N.~W., {Hewett}, P., {et~al.} 2004, VizieR Online Data
  Catalog, J/MNRAS/351/1071

\bibitem[Astropy Collaboration et al.(2013)]{2013A&A...558A..33A} Astropy Collaboration, Robitaille, T.~P., Tollerud, E.~J., et al.\ 2013, \aap, 558, A33 

\bibitem[{{Azimlu} {et~al.}(2011){Azimlu}, {Marciniak}, \& {Barmby}}]{Azimlu}
{Azimlu}, M., {Marciniak}, R., \& {Barmby}, P. 2011, \aj, 142, 139,

\bibitem[Bertin \& Arnouts(1996)]{1996A&AS..117..393B} Bertin, E., \& Arnouts, S.\ 1996, \aaps, 117, 393 

\bibitem[{{Bonanos} {et~al.}(2019){Bonanos}, {Yang}, {Sokolovsky}, {Gavras},
  {Hatzidimitriou}, {Bellas-Velidis}, {Kakaletris}, {Lennon}, {Nota}, {White},
  {Whitmore}, {Anastasiou}, {Arevalo}, {Arviset}, {Baines}, {Budavari},
  {Charmand aris}, {Chatzichristodoulou}, {Dimas}, {Duran}, {Georgantopoulos},
  {Karampelas}, {Laskaris}, {Lianou}, {Livanis}, {Lubow}, {Manouras},
  {Moretti}, {Paraskeva}, {Pouliasis}, {Rest}, {Salgado}, {Sonnentrucker},
  {Spetsieri}, {Taylor}, \& {Tsinganos}}]{Bonanos}
{Bonanos}, A.~Z., {Yang}, M., {Sokolovsky}, K.~V., {et~al.} 2019, VizieR Online
  Data Catalog, J/A+A/630/A92
  
\bibitem[Corrales(2015)]{2015ApJ...805...23C} Corrales, L.\ 2015, \apj, 805, 23

\bibitem[Ferland et al.(2013)]{2013RMxAA..49..137F} Ferland, G.~J., Porter, R.~L., van Hoof, P.~A.~M., et al.\ 2013, \rmxaa, 49, 137

\bibitem[{{Fliri} {et~al.}(2006){Fliri}, {Riffeser}, {Seitz}, \&
  {Bender}}]{Fliri}
{Fliri}, J., {Riffeser}, A., {Seitz}, S., \& {Bender}, R. 2006, VizieR Online
  Data Catalog, J/A+A/445/423
  
\bibitem[Gaia Collaboration et al.(2016)]{2016A&A...595A...1G} Gaia Collaboration, Prusti, T., de Bruijne, J.~H.~J., et al.\ 2016, \aap, 595, A1. doi:10.1051/0004-6361/201629272

\bibitem[Gaia Collaboration et al.(2018)]{2018A&A...616A...1G} Gaia Collaboration, Brown, A.~G.~A., Vallenari, A., et al.\ 2018, \aap, 616, A1. doi:10.1051/0004-6361/201833051

\bibitem[{{Gaia Collaboration}(2018)}]{GaiaDR2}
{Gaia Collaboration}. 2018, VizieR Online Data Catalog, I/345

\bibitem[{{Halliday} {et~al.}(2006){Halliday}, {Carter}, {Bridges}, {Jackson},
  {Wilkinson}, {Quinn}, {Evans}, {Douglas}, {Merrett}, {Merrifield},
  {Romanowsky}, {Kuijken}, \& {Irwin}}]{Halliday}
{Halliday}, C., {Carter}, D., {Bridges}, T.~J., {et~al.} 2006, VizieR Online
  Data Catalog, J/MNRAS/369/97

\bibitem[Hanisch \& Biemesderfer(1989)]{1989BAAS...21..780H} Hanisch, R.~J., \& Biemesderfer, C.~D.\ 1989, \baas, 21, 780 

\bibitem[{{Heinze} {et~al.}(2018){Heinze}, {Tonry}, {Denneau}, {Flewelling},
  {Stalder}, {Rest}, {Smith}, {Smartt}, \& {Weiland}}]{Heinze}
{Heinze}, A.~N., {Tonry}, J.~L., {Denneau}, L., {et~al.} 2018, \aj, 156, 241


\bibitem[{{Hornoch} \& {Kucakova}(2016{\natexlab{a}})}]{HornochAug}
{Hornoch}, K., \& {Kucakova}, H. 2016{\natexlab{a}}, The Astronomer's Telegram,
  9373, 1

\bibitem[{{Hornoch} \& {Kucakova}(2016{\natexlab{b}})}]{HornochMar}
---. 2016{\natexlab{b}}, The Astronomer's Telegram, 8785, 1

\bibitem[{{Hornoch} \& {Kucakova}(2019{\natexlab{a}})}]{HornochOct}
---. 2019{\natexlab{a}}, The Astronomer's Telegram, 13207, 1

\bibitem[{{Hornoch} \& {Kucakova}(2019{\natexlab{b}})}]{HornochJul}
---. 2019{\natexlab{b}}, The Astronomer's Telegram, 12925, 1

\bibitem[{{Humphreys} {et~al.}(2017{\natexlab{a}}){Humphreys}, {Davidson},
  {Hahn}, {Martin}, \& {Weis}}]{HumphreysV}
{Humphreys}, R.~M., {Davidson}, K., {Hahn}, D., {Martin}, J.~C., \& {Weis}, K.
  2017{\natexlab{a}}, VizieR Online Data Catalog, J/ApJ/844/40

\bibitem[{{Humphreys} {et~al.}(2017{\natexlab{b}}){Humphreys}, {Gordon},
  {Martin}, {Weis}, \& {Hahn}}]{HumphreysIV}
{Humphreys}, R.~M., {Gordon}, M.~S., {Martin}, J.~C., {Weis}, K., \& {Hahn}, D.
  2017{\natexlab{b}}, VizieR Online Data Catalog, J/ApJ/836/64

\bibitem[{{Johnson} {et~al.}(2015){Johnson}, {Seth}, {Dalcanton}, {Wallace},
  {Simpson}, {Lintott}, {Kapadia}, {Skillman}, {Caldwell}, {Fouesneau},
  {Weisz}, {Williams}, {Beerman}, {Gouliermis}, \& {Sarajedini}}]{Johnson}
{Johnson}, L.~C., {Seth}, A.~C., {Dalcanton}, J.~J., {et~al.} 2015, \apj, 802,
  127
  

\bibitem[{{Kim} {et~al.}(2007){Kim}, {Lee}, {Park}, {Hwang}, {Geisler},
  {Sarajedini}, {Harris}, \& {Seguel}}]{Kim}
{Kim}, S.~C., {Lee}, M.~G., {Park}, H.~S., {et~al.} 2007, in Astronomical
  Society of the Pacific Conference Series, Vol. 362, The Seventh Pacific Rim
  Conference on Stellar Astrophysics, ed. Y.~W. {Kang}, H.~W. {Lee}, K.~C.
  {Leung}, \& K.~S. {Cheng}, 286

\bibitem[{{Kodric} {et~al.}(2018){Kodric}, {Riffeser}, {Hopp}, {Goessl},
  {Seitz}, {Bender}, {Koppenhoefer}, {Obermeier}, {Snigula}, {Lee}, {Burgett},
  {Draper}, {Hodapp}, {Kaiser}, {Kudritzki}, {Metcalfe}, {Tonry}, \&
  {Wainscoat}}]{Kodric}
{Kodric}, M., {Riffeser}, A., {Hopp}, U., {et~al.} 2018, VizieR Online Data
  Catalog, J/AJ/156/130

\bibitem[{{Kurtev}(2003)}]{Kurtev}
{Kurtev}, R.~G. 2003, Astronomische Nachrichten, 324, 265

\bibitem[Leahy et al.(2018)]{2018AJ....156..269L} Leahy, D.~A., Bianchi, L., \& Postma, J.~E.\ 2018, \aj, 156, 269

\bibitem[Leahy et al.(2020)]{2020ApJS..247...47L} Leahy, D.~A., Postma, J., Chen, Y., et al.\ 2020, \apjs, 247, 47. doi:10.3847/1538-4365/ab77a9


\bibitem[Leahy \& Chen(2020)]{2020ApJS..250...23L} Leahy, D.~A. \& Chen, Y.\ 2020, \apjs, 250, 23

\bibitem[Leahy et al.(2020b)]{2020arXiv201202727L} Leahy, D.~A., Postma, J., Buick, M., et al.\ 2020, arXiv:2012.02727

\bibitem[Martin et al.(2005)]{2005ApJ...619L...1M} Martin, D.~C., Fanson, J., Schiminovich, D., et al.\ 2005, \apjl, 619, L1. doi:10.1086/426387

\bibitem[McConnachie et al.(2005)]{2005MNRAS.356..979M} McConnachie, A.~W., Irwin, M.~J., Ferguson, A.~M.~N., et al.\ 2005, \mnras, 356, 979. doi:10.1111/j.1365-2966.2004.08514.x


\bibitem[{{Neugent} {et~al.}(2012){Neugent}, {Massey}, \& {Georgy}}]{Neugent}
{Neugent}, K.~F., {Massey}, P., \& {Georgy}, C. 2012, \apj, 759, 11

\bibitem[Lamport(1994)]{lamport94} Lamport, L. 1994, LaTeX: A Document Preparation System, 2nd Edition (Boston, Addison-Wesley Professional)

\bibitem[Ochsenbein et al.(2000)]{2000A&AS..143...23O} Ochsenbein, F., Bauer, P., \& Marcout, J.\ 2000, \aaps, 143, 23. doi:10.1051/aas:2000169

\bibitem[{{Ovcharov} {et~al.}(2015){Ovcharov}, {Kostov}, {Kurtenkov},
  {Valcheva}, \& {Nedialkov}}]{Ovcharov}
{Ovcharov}, E., {Kostov}, A., {Kurtenkov}, A., {Valcheva}, A., \& {Nedialkov},
  P. 2015, The Astronomer's Telegram, 7065, 1

\bibitem[{{Peacock} {et~al.}(2010){Peacock}, {Maccarone}, {Knigge}, {Kundu},
  {Waters}, {Zepf}, \& {Zurek}}]{Peacock}
{Peacock}, M.~B., {Maccarone}, T.~J., {Knigge}, C., {et~al.} 2010, VizieR
  Online Data Catalog, J/MNRAS/402/803
  
  \bibitem[Postma \& Leahy(2017)]{2017PASP..129k5002P} Postma, J.~E. \& Leahy, D.\ 2017, \pasp, 129, 115002. doi:10.1088/1538-3873/aa8800
  
  \bibitem[Postma \& Leahy(2020)]{2020PASP..132e4503P} Postma, J.~E. \& Leahy, D.\ 2020, \pasp, 132, 054503. doi:10.1088/1538-3873/ab7ee8

\bibitem[{{Samus'} {et~al.}(2017){Samus'}, {Kazarovets}, {Durlevich},
  {Kireeva}, \& {Pastukhova}}]{Samus}
{Samus'}, N.~N., {Kazarovets}, E.~V., {Durlevich}, O.~V., {Kireeva}, N.~N., \&
  {Pastukhova}, E.~N. 2017, Astronomy Reports, 61, 80

\bibitem[Schwarz et al.(2011)]{2011ApJS..197...31S} Schwarz, G.~J., Ness, J.-U., Osborne, J.~P., et al.\ 2011, \apjs, 197, 31  
\bibitem[Vogt et al.(2014)]{2014ApJ...793..127V} Vogt, F.~P.~A., Dopita, M.~A., Kewley, L.~J., et al.\ 2014, \apj, 793, 127  

 \bibitem[Singh et al.(2014)]{2014SPIE.9144E..1SS} Singh, K.~P., Tandon, S.~N., Agrawal, P.~C., et al.\ 2014, \procspie, 9144, 91441S. doi:10.1117/12.2062667


\bibitem[{{Soraisam} {et~al.}(2019){Soraisam}, {Lee}, {Narayan}, {Matheson}, \&
  {Saha}}]{Soraisam}
{Soraisam}, M., {Lee}, C.-H., {Narayan}, G., {Matheson}, T., \& {Saha}, A.
  2019, The Astronomer's Telegram, 13210, 1

\bibitem[Tandon et al.(2017)]{2017AJ....154..128T} Tandon, S.~N., Subramaniam, A., Girish, V., et al.\ 2017, \aj, 154, 128

\bibitem[{{Tandon} {et~al.}(2020){Tandon}, {Postma}, {Joseph}, {Devaraj},
  {Subramaniam}, {Barve}, {George}, {Ghosh}, {Girish}, {Hutchings}, {Kamath},
  {Kathiravan}, {Kumar}, {Lancelot}, {Leahy}, {Mahesh}, {Mohan},
  {Nagabhushana}, {Pati}, {Rao}, {Sankarasubramanian}, {Sriram}, \&
  {Stalin}}]{Tandon}
{Tandon}, S.~N., {Postma}, J., {Joseph}, P., {et~al.} 2020, \aj, 159, 158

\bibitem[{{Taylor}(2005)}]{2005ASPC..347...29T}
{Taylor}, M.~B. 2005, in Astronomical Society of the Pacific Conference Series,
  Vol. 347, Astronomical Data Analysis Software and Systems XIV, ed.
  P.~{Shopbell}, M.~{Britton}, \& R.~{Ebert}, 29

\bibitem[{{Vilardell} {et~al.}(2006){Vilardell}, {Ribas}, \&
  {Jordi}}]{Vilardell}
{Vilardell}, F., {Ribas}, I., \& {Jordi}, C. 2006, VizieR Online Data Catalog,
  J/A+A/459/321

\bibitem[{{Walterbos} \& {Braun}(1994)}]{Walterbos}
{Walterbos}, R.~A.~M., \& {Braun}, R. 1994, VizieR Online Data Catalog,
  J/A+AS/92/625

\bibitem[{{Williams} {et~al.}(2014){Williams}, {Lang}, {Dalcanton}, {Dolphin},
  {Weisz}, {Bell}, {Bianchi}, {Byler}, {Gilbert}, {Girardi}, {Gordon},
  {Gregersen}, {Johnson}, {Kalirai}, {Lauer}, {Monachesi}, {Rosenfield},
  {Seth}, \& {Skillman}}]{PHAT}
{Williams}, B.~F., {Lang}, D., {Dalcanton}, J.~J., {et~al.} 2014, \apjs, 215,
  9

\bibitem[{{Williams} \& {Darnley}(2015)}]{Williams2015}
{Williams}, S.~C., \& {Darnley}, M.~J. 2015, The Astronomer's Telegram, 8218, 1


\end{thebibliography}
\end{document}